\documentclass[12pt]{article}

\usepackage{amssymb,bm}
\usepackage{amsmath}
\usepackage{amsthm}
\usepackage{graphicx}
\usepackage[active]{srcltx} 
\usepackage{hyperref} 
\hypersetup{pdfborder=0 0 0}
\newtheorem{theorem}{{\sc Theorem}}[section]

\newtheorem{definition}[theorem]{Definition}

\newcommand{\bb}[1]{\mathbb{ #1}}

\bmdefine\Bone{1}

\newcommand{\dev}[1]{\mathrm{dev}(#1)}

\newcommand{\bra}[1]{\overline{#1}}

\newcommand{\Trc}{\mathrm{Tr}\,}

\newcommand{\cof}{\mathrm{cof}}

\newcommand{\tns}[1]{#1\otimes #1}
\newcommand{\hf}{\displaystyle\frac{1}{2}}
\newcommand{\nth}[1]{\displaystyle\frac{1}{#1}}

\newcommand{\dif}[2]{\displaystyle\frac{\partial #1}{\partial #2}}
\newcommand{\Grad}{\nabla}
\newcommand{\Div}{\nabla \cdot}

\newcommand{\Md}{\partial}

\newcommand{\Tld}[1]{\widetilde{#1}}
\newcommand{\hess}[2]{\displaystyle\frac{\partial^2 #1}{\partial #2^2}}
\newcommand{\mix}[3]{\displaystyle\frac{\partial^2 #1}{\partial #2\partial #3}}
\newcommand{\av}[1]{\langle #1 \rangle}

\def\XXint#1#2#3{{\setbox0=\hbox{$#1{#2#3}{\int}$ }
\vcenter{\hbox{$#2#3$ }}\kern-.6\wd0}}

\newcommand{\jump}[1]{\lbrack\!\lbrack #1 \rbrack\!\rbrack}
\newcommand{\lump}[1]{\lbrace\skew{-14.7}\lbrace\!\!#1\!\!\skew{14.7}\rbrace\rbrace}

\newcommand{\mat}[4]{\left[\begin{array}{cc}
\displaystyle{#1}&\displaystyle{#2}\\[1ex]
\displaystyle{#3}&\displaystyle{#4}\end{array}\right]}

\newcommand{\bc}{boundary condition}

\newcommand{\rhs}{right-hand side}

\newcommand{\mc}{microstructure}

\newcommand{\WLOG}{without loss of generality}
\newcommand{\nbh}{neighborhood}
\newcommand{\IFF}{if and only if }

\newcommand{\Ga}{\alpha}
\newcommand{\Gb}{\beta}
\newcommand{\Gd}{\delta}

\newcommand{\eps}{\varepsilon}
\newcommand{\Gve}{\varepsilon}

\newcommand{\Gg}{\gamma}

\newcommand{\Gl}{\lambda}

\newcommand{\Gth}{\theta}

\newcommand{\Go}{\omega}

\newcommand{\GD}{\Delta}

\newcommand{\GO}{\Omega}

\bmdefine\BGa{\alpha}
\bmdefine\BGb{\beta}
\bmdefine\BGd{\delta}
\bmdefine\BGe{\epsilon}
\bmdefine\BGve{\varepsilon}
\bmdefine\BGf{\phi}
\bmdefine\BGvf{\varphi}
\bmdefine\BGg{\gamma}
\bmdefine\BGc{\chi}
\bmdefine\BGi{\iota}
\bmdefine\BGk{\kappa}
\bmdefine\BGl{\lambda}
\bmdefine\BGn{\eta}
\bmdefine\BGm{\mu}
\bmdefine\BGv{\nu}
\bmdefine\BGp{\pi}
\bmdefine\BGth{\theta}
\bmdefine\BGvth{\vartheta}
\bmdefine\BGr{\rho}
\bmdefine\BGvr{\varrho}
\bmdefine\BGs{\sigma}
\bmdefine\BGvs{\varsigma}
\bmdefine\BGt{\tau}
\bmdefine\BGj{\tau}
\bmdefine\BGu{\upsilon}
\bmdefine\BGo{\omega}
\bmdefine\BGx{\xi}
\bmdefine\BGy{\psi}
\bmdefine\BGz{\zeta}
\bmdefine\BGD{\Delta}
\bmdefine\BGF{\Phi}
\bmdefine\BGG{\Gamma}
\bmdefine\BGL{\Lambda}
\bmdefine\BGP{\Pi}
\bmdefine\BGT{\Theta}
\bmdefine\BGS{\Sigma}
\bmdefine\BGU{\Upsilon}
\bmdefine\BGO{\Omega}
\bmdefine\BGX{\Xi}
\bmdefine\BGY{\Psi}

\bmdefine\BFM{\mathfrak{M}}
\bmdefine\BFb{\mathfrak{b}}
\bmdefine\BFk{\mathfrak{k}}
\bmdefine\BFm{\mathfrak{m}}
\bmdefine\BFu{\mathfrak{u}}
\bmdefine\BFv{\mathfrak{v}}

\newcommand{\CS}{{\mathcal S}}

\bmdefine\BCA{{\mathcal A}}
\bmdefine\BCB{{\mathcal B}}
\bmdefine\BCC{{\mathcal C}}
\bmdefine\BCD{{\mathcal D}}
\bmdefine\BCE{{\mathcal E}}
\bmdefine\BCF{{\mathcal F}}
\bmdefine\BCG{{\mathcal G}}
\bmdefine\BCH{{\mathcal H}}
\bmdefine\BCI{{\mathcal I}}
\bmdefine\BCJ{{\mathcal J}}
\bmdefine\BCK{{\mathcal K}}
\bmdefine\BCL{{\mathcal L}}
\bmdefine\BCM{{\mathcal M}}
\bmdefine\BCN{{\mathcal N}}
\bmdefine\BCO{{\mathcal O}}
\bmdefine\BCP{{\mathcal P}}
\bmdefine\BCQ{{\mathcal Q}}
\bmdefine\BCR{{\mathcal R}}
\bmdefine\BCS{{\mathcal S}}
\bmdefine\BCT{{\mathcal T}}
\bmdefine\BCU{{\mathcal U}}
\bmdefine\BCV{{\mathcal V}}
\bmdefine\BCW{{\mathcal W}}
\bmdefine\BCX{{\mathcal X}}
\bmdefine\BCY{{\mathcal Y}}
\bmdefine\BCZ{{\mathcal Z}}

\bmdefine\Bzr{ 0}
\bmdefine\Ba{ a}
\bmdefine\Bb{ b}
\bmdefine\Bc{ c}
\bmdefine\Bd{ d}
\bmdefine\Be{ e}
\bmdefine\Bf{ f}
\bmdefine\Bg{ g}
\bmdefine\Bh{ h}
\bmdefine\Bi{ i}
\bmdefine\Bj{ j}
\bmdefine\Bk{ k}
\bmdefine\Bl{ l}
\bmdefine\Bm{ m}
\bmdefine\Bn{ n}
\bmdefine\Bo{ o}
\bmdefine\Bp{ p}
\bmdefine\Bq{ q}
\bmdefine\Br{ r}
\bmdefine\Bs{ s}
\bmdefine\Bt{ t}
\bmdefine\Bu{ u}
\bmdefine\Bv{ v}
\bmdefine\Bw{ w}
\bmdefine\Bx{ x}
\bmdefine\By{ y}
\bmdefine\Bz{ z}
\bmdefine\BA{ A}
\bmdefine\BB{ B}
\bmdefine\BC{ C}
\bmdefine\BD{ D}
\bmdefine\BE{ E}
\bmdefine\BF{ F}
\bmdefine\BG{ G}
\bmdefine\BH{ H}
\bmdefine\BI{ I}
\bmdefine\BJ{ J}
\bmdefine\BK{ K}
\bmdefine\BL{ L}
\bmdefine\BM{ M}
\bmdefine\BN{ N}
\bmdefine\BO{ O}
\bmdefine\BP{ P}
\bmdefine\BQ{ Q}
\bmdefine\BR{ R}
\bmdefine\BS{ S}
\bmdefine\BT{ T}
\bmdefine\BU{ U}
\bmdefine\BV{ V}
\bmdefine\BW{ W}
\bmdefine\BX{ X}
\bmdefine\BY{ Y}
\bmdefine\BZ{ Z}

\usepackage{color}
\title{Solid phase transitions in the liquid limit}
\author{Yury Grabovsky\thanks{Department of Mathematics, Temple University, Philadelphia, PA 19122, USA} \and Lev Truskinovsky\thanks{PMMH, CNRS -- UMR 7636, ESPCI, PSL,  75005 Paris, France}}
\begin{document}
\maketitle

\hspace{44ex}{\footnotesize Dedicated to the memory of Jerry Ericksen, a wizard}

\begin{abstract}
  We address the fundamental difference between   solid-solid  and liquid-liquid phase transitions   within the Ericksen's  nonlinear elasticity paradigm.  To highlight ideas, we consider   the simplest nontrivial  2D problem and work with a prototypical  two-phase  Hadamard material which allows one to weaken the  rigidity  and  explore  the nature of solid-solid phase transitions in    a  ``near-liquid'' limit. In the language of calculus of variations we probe limits of quasiconvexity in an ``almost liquid'' solid by comparing the thresholds for  cooperative (laminate based)  and non-cooperative (inclusion based)  nucleation.  Using these two types of nucleation tests we  obtain  for our model material surprisingly tight two-sided  bounds on the elastic binodal without directly computing the quasiconvex envelope.  
\end{abstract}
\section{Introduction}
\setcounter{equation}{0}
\label{sec:intro}
In 1975 J. Ericksen placed for the first time the problem of   solid-solid phase transitions 
in  the framework of nonlinear elasticity theory. He effectively reformulated the classical
physics problem  as a problem of vectorial calculus of variations. The
contemporaneous physical theory viewed non-hydrostatically stressed solids as
metastable and therefore did not distinguish between solid-solid and
liquid-liquid phase transitions. Ericksen's insight   that during phase transitions the non-hydrostatic stresses may  persist at time scales of interest,  revolutionized elasticity theory  and initiated the extremely successful research program of studying materials with
non-quasiconvex energies \cite{silh13,Dak08,krro19,ball02,ciar21}.
The goal of the present  paper is to elucidate the difference between solid-solid and
liquid-liquid phase transitions within the Ericksen's nonlinear elasticity
paradigm.

From the perspective of elasticity theory, the main difference between liquids
and solids is that liquids do not resist shear \cite{chlw95,dcbunv16}. This
degeneracy in the elastic constitutive structure of liquids is responsible for
their peculiar behavior during first order phase transitions vis a vis the
behavior of solids, characterized by finite rigidity \cite{golu89}.  
For  both solids and liquids, reaching phase equilibrium often means the formation of
phase mixtures. However, while the phase
organization in liquid phase transitions is largely controlled by surface
tension, in solid phase transitions the dominance of elastic long-range
interactions leaves to surface tension only a minor role of a scale
selection. The  dependence of energy  on the geometric arrangement of
the phases  in solids  leads to specific morphologies,   largely controlled by
the interplay between  the  location of the energy wells  and their  kinematic
compatibility \cite{baja15}.

First order phase transitions in liquids are well understood at both
physical and mathematical levels \cite{lali13v5,Dak08}. The reason is that   the scalar problem confronted in the liquid case is fully solvable \cite{daco81}. Instead, despite many dedicated efforts, largely inspired by the pioneering
contributions of  Ericksen himself
\cite{erick75,Ericksen:1980:SPT,erick87,Ericksen:1991:KCC,erick92a},  the
mathematical understanding  of elastic phase transitions in solids is still
far from being complete. In particular,   some basic  underlying   vectorial
problems of the calculus of variations remain unsolved \cite{ball02,ball10}.  

To set the stage, we recall that in nonlinear elasticity  the  energy functional  can be written in the form 
$
  E[\By]=\int_{\GO}W( \Grad\By)d\Bx,
$
where $W(\BF)$ is the energy density function describing the elastic
properties of the material.
 For the energy minimizing
 configurations  the physically informed energy density $W(\BF)$
 is replaced   by its quasiconvexification  $QW(\BF)$ \cite{daco82}.  The latter can be given by an implicit  formula which becomes  explicit only 
 if one   knows the energy minimizing microstructures.  In the case of liquids  the geometry of such microstructures is irrelevant and the construction  of  $QW(\BF)$ reduces to convexification. In the case of solids, the task of finding the  equilibrium microstructures in a generic setting is daunting.

With the aim of  building a bridge between elastic phase transitions in
liquids and solids, we consider in this paper a special limit of ``near-liquid''
solids which are characterized  by an arbitrarily weak resistance to
shear. We pose  the general question of how in such a limit the  tight
control on the geometry of optimal microstructures by elastic  interactions is
lost. To answer this question
we address a simpler problem of describing in this limit the boundary of the
set of stable single-phase configurations.
Such  problem can  be  solved in the case of ``strongly-solid'' elastic phase
transitions when the equilibrium microstructures are   simple laminates
\cite{grtrsolid}. The goal of  the present paper is to understand  the opposite,
``weakly-solid''  limit, when the simplest laminate-based microstructures are clearly  suboptimal.

In physics literature the boundary of the
set of stable single-phase configurations is delineated by the classical Maxwell-Gibbs
 equilibrium conditions  which were originally developed to describe  phase equilibrium  in liquids \cite{maxwell1875, gibbs1876}. These conditions allow one to identify the
 homogeneous  configurations that are unstable to perturbations that are small
 in extent but not necessarily in magnitude. Such configurations  are  known in physics as constituting the  binodal region \cite{vandW03}. In the
mathematical theory of elastic phase transitions,  the ``solid''  analog of the  binodal
region  would incorporate the  homogeneous states that fail to be  strong
local minima of the energy functional. 

In this perspective,  the  binodal region   is a
subset in the configuration space  of strain measures where the quasiconvex
envelope lies below the  energy density. Locating the  boundary of the
binodal region (called the \emph{binodal} in physics) in the general setting 
constitutes a major challenge.  While remaining nontrivial,
this task appears, at least a priori, as more tractable than the task of
constructing the  actual quasiconvex envelope. In  the   present paper we
address the problem of the binodal in the ``near-liquid'' limit and  show  that   in such a limit  the explicit knowledge of the binodal  can  lead directly to explicit formulas for at least some parts of the  quasiconvex envelope.

 In our prior work we have developed general methods
for identifying subsets of the binodal marking the emergence (nucleation) of coherent
regions of a new phase \cite{grtrmms}. The main tool in such  analysis is the
characterization of the jump set \cite{grtrpe}---a codimension one variety in the phase
space corresponding to nucleation of coherent laminae. The knowledge
  of the jump set  provides a general way to  constrain (bound) the  binodal
  from within, without any guarantee that the bound is optimal. Applying our
  methods of identifying both stable and unstable parts of the jump set
  ultimately leads to a  realistic approximation  of the
  whole  binodal in the ``near-liquid''  limit.

\begin{figure}[t]
  \centering
  \includegraphics[scale=0.3]{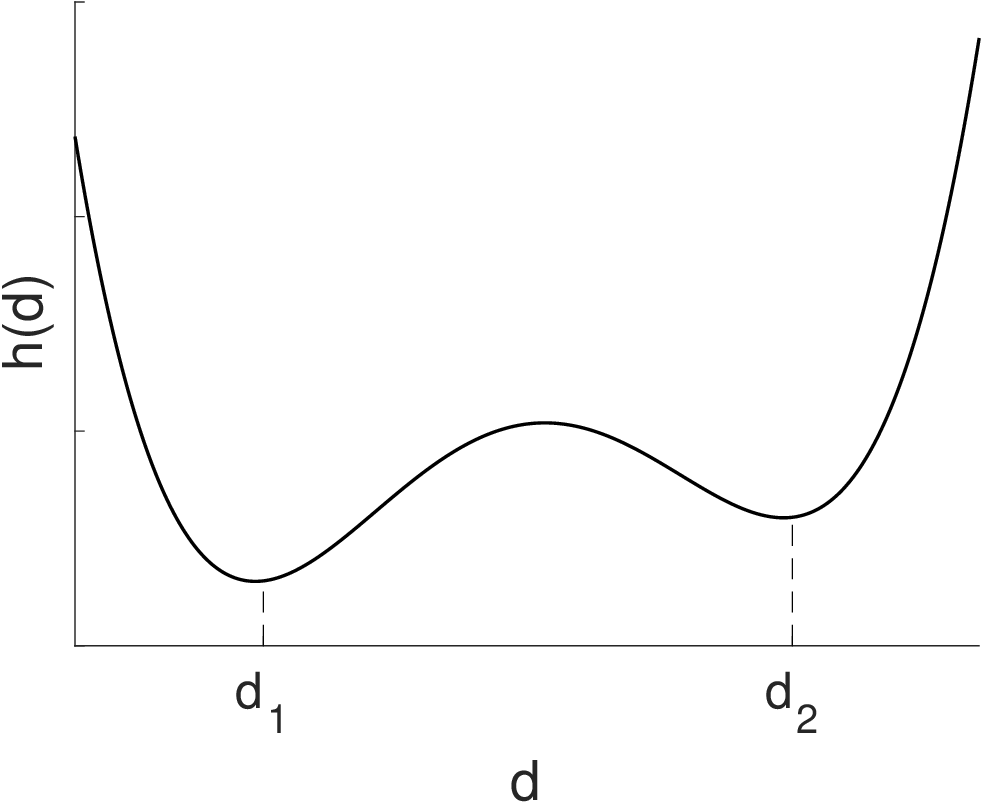}
  \caption{Double-well structure  of the energy density $h(d)$.}
  \label{fig:dblwell}
\end{figure}

To highlight  ideas, we focus in this paper  on the simplest  family of non-quasiconvex energy densities   describing   Hadamard materials \cite{hada03,john66}:
$
  W(\BF)=\frac{\mu}{2}|\BF|^{2}+h(\det\BF).
$ 
Specifically, we are interested in the case of  two space
dimensions\footnote{ In principle, our methodology is 
    also applicable in 3D. In this paper we have chosen a 2D setting to make the ideas
    and techniques fully transparent  and  to be able to illustrate them
    graphically.}
    and assume that the function $h(d)$ describes  a generic 
double-well potential  modeling  
isotropic-to-isotropic  phase transitions (see Fig.~\ref{fig:dblwell}).
The main advantage of  this class of energy densities is that it
contains  a  single parameter $\mu$ which can be viewed as a scale of  the  effective rigidity.
By varying this scale  one  can study the entire range of intermediate rigidity responses from ``strong'' ($\mu \gg 1$) to ``weak'' ($\mu  \ll 1$) and in this way expose the crossover from ``solid'' to  ``liquid''  behavior.

An interesting property of Hadamard materials is the subtle \emph{asymmetry} between  the two isotropic phases. It is intriguing, as it can be attributed
neither to the difference between the elastic moduli of the phases nor to the
geometric nonlinearity of the model. Indeed, it persists even if  the     two   wells of the nonlinear potential  $h(d)$ are identical  in the sense that they have   the same height and the same tangential elastic moduli. The only difference between the two phases
is then governed by the small term $\mu|\BF|^{2}/2$, that is larger at the low
density phase and smaller at the high density phase. However, due to this difference, and in contrast to  the case of an ostensibly similar  geometrically  linearly    model \cite{khach67,pipk91,DW},   
 the two phases of an  Hadamard material admit rather dissimilar ecosystems of instability mechanisms. 
Thus, rather remarkably, the instability of the low energy phase proceeds by
nucleation of a compact region of the high energy minority phase, in contrast to the
formation of laminar \mc, when the high energy phase loses its stability. We show that such an   asymmetry  leads to a   coexistence of ``strongly-solid'' and ``weakly-solid'' (or even ``quasi-liquid'') responses inside a single material model. In particular,  even in the absence of   hysteresis,   the   direct and  reverse   phase transitions   in such material can  proceed  according to morphologically distinct transformation mechanisms.

To understand this behavior, we first recall that while for an Hadamard material the double well energy structure is described
in terms of a single scalar  potential  $h(d)$,  the results of relaxation of
$W(\BF)$ are nontrivial due to the inherent incompatibility  of the energy
wells associated with  the purely volumetric nature of the implied phase transition.
 In the case of only one well, when the possibility of phase transition is absent,    the result of such relaxation  is trivial as it is known that $W(\BF)$ is quasiconvex \IFF $h(d)$ is
convex \cite{bamu84}. The relaxation of $W(\BF)$ with two wells and non-convex $h(d)$ is known   for the ``infinitely-weak'' solids (effectively liquids)  with $\mu=0$,
where $QW(\BF)=h^{**}(\det\BF)$  \cite{daco81} and 
in the ``strongly-solid'' limit when the shear modulus $\mu$ is sufficiently
large \cite{grtrsolid}. In the latter case the quadratic term in the energy
dominates the double-well term and the formula for $QW(\BF)$ couples $|\BF|$
and $\det\BF$ placing the relaxed energy between $W(\BF)$ above and
$U(\BF)=\frac{\mu}{2}|\BF|^{2}+h^{**}(\det\BF)$ below.
\begin{figure}[t]
  \centering
\includegraphics[scale=0.32]{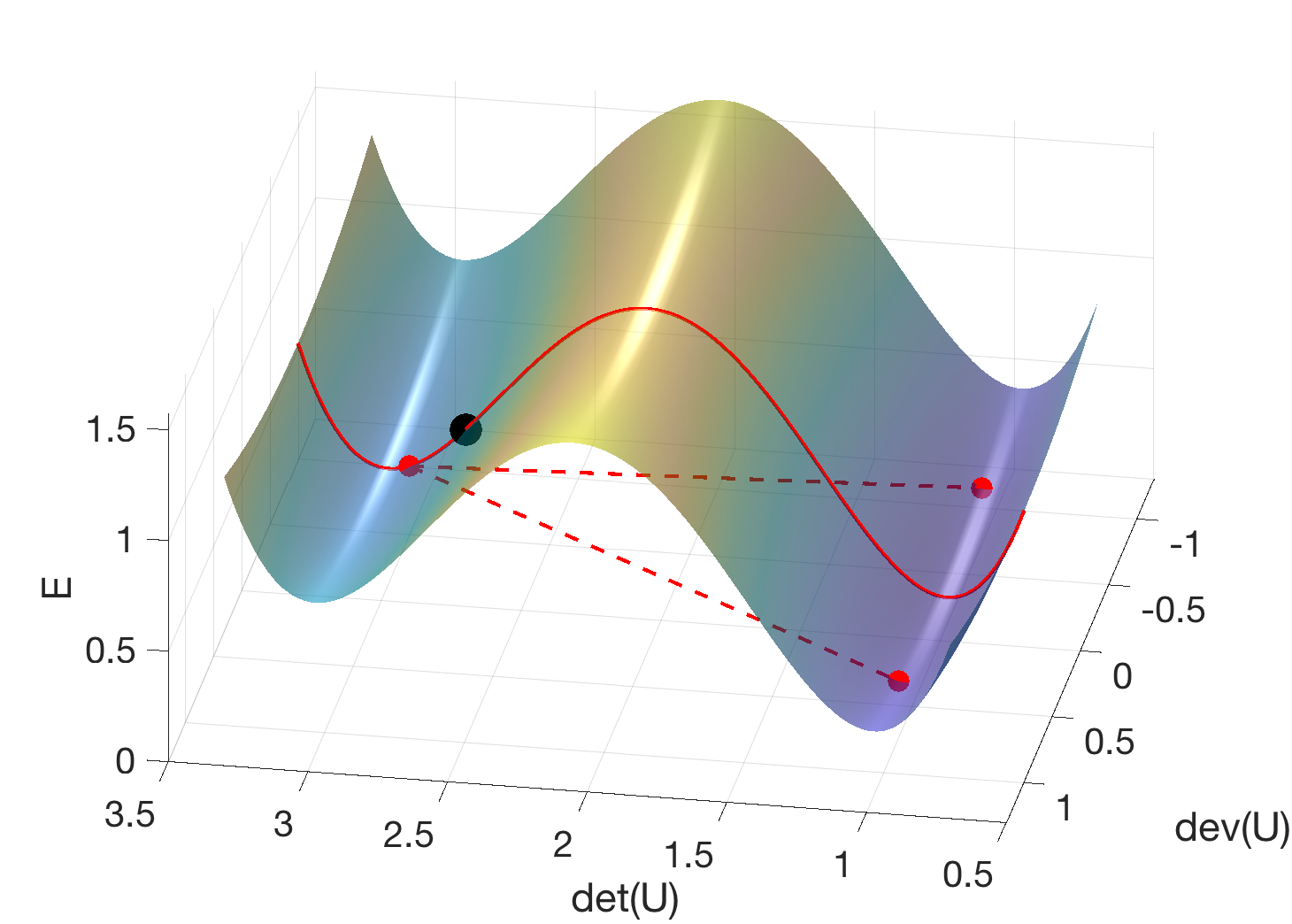}\hspace{3ex}
\includegraphics[scale=0.32]{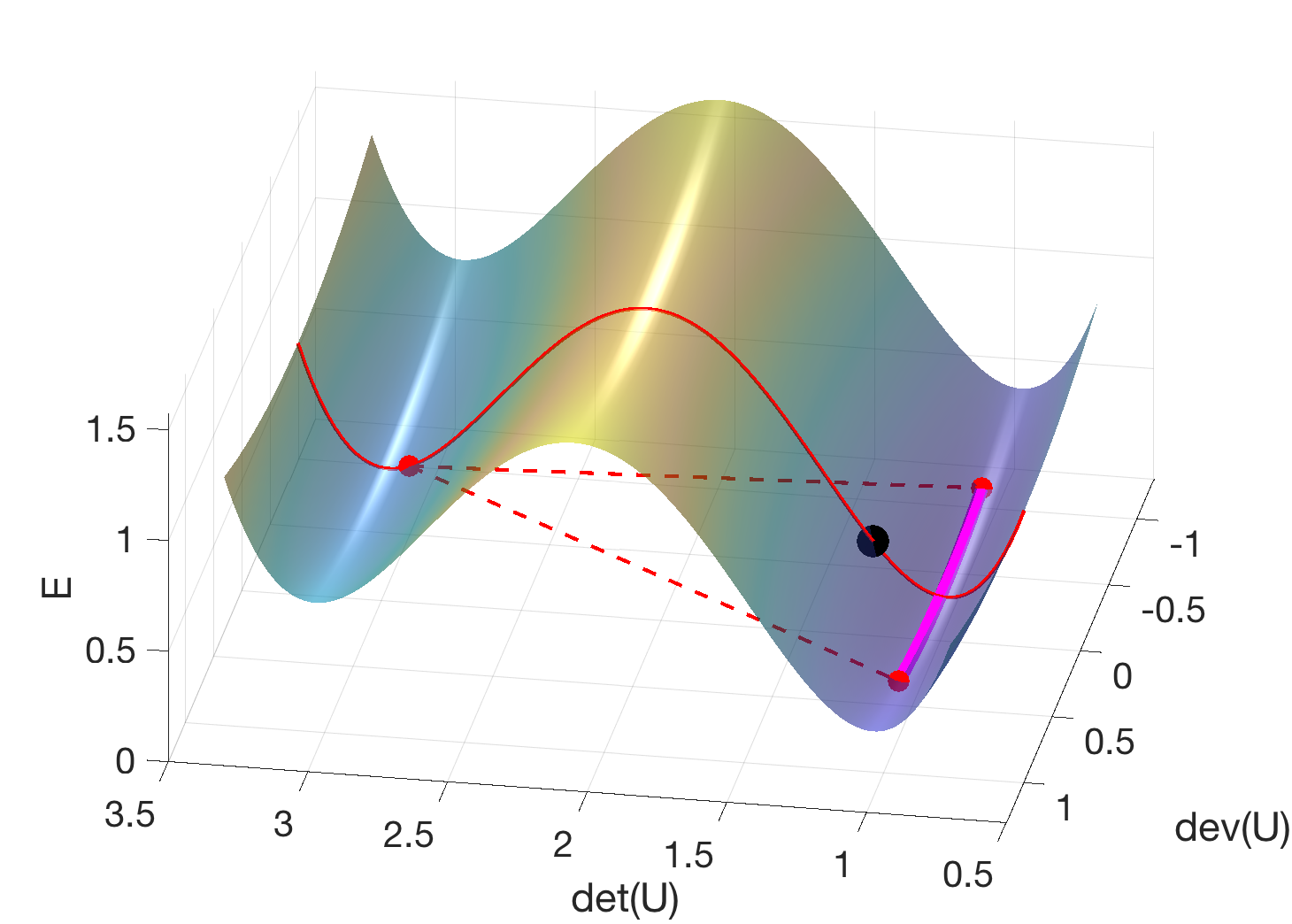}
\caption{The schematic energy landscape of the Hadamard material showing  the elastic fields at the point of instability of the homogeneous state (incipient phase transition). Left panel: instability of the high energy
  phase. Right panel: instability of the low energy
  phase. The initial state is represented by a black dot.}
\label{fig:energy}
\end{figure}

In the present  paper we show that the lower bound on  the rigidity measure $\mu$ in \cite{grtrsolid}
was not a technical limitation, and that, as $\mu$ decreases,  the above  ``strongly-solid''
expression  for $QW(\BF)$ ceases to be valid in the subsets of the binodal
corresponding to hydrostatic Dirichlet \bc s.
More specifically, we show that in the limit of small $\mu$, the relaxation of
$W(\BF)$ goes through a chain of structural transitions with simple lamination
persisting either only for high shear hard device loading and in the vicinity
of the higher energy well.  Close to the low energy well it is replaced by
more complex optimal microstructures which remain not fully
characterized. Fig.~\ref{fig:energy} shows the schematic energy landscape of
the Hadamard material in the variables $\det\BU=\Gve_{1}\Gve_{2}$, and
$\dev{\BU}=\Gve_{1}-\Gve_{2}$, where $\BU=(\BF^{T}\BF)^{1/2}$ and $\Gve_{1}$
and $\Gve_{2}$ are the singular values of the deformation gradient $\BF$; in
both panels the black dot represents the applied affine hydrostatic Dirichlet
boundary conditions.  The left panel shows the fields and their rank-one
connections at the onset of the instability of the high energy
phase. The compatible field (inside a simple laminate) takes two
  values represented by the red dot on the left and one of the two red dots on
  the right. The right panel shows the fields at the onset of the instability
  in the low energy phase. The corresponding compatible field
takes values in the set represented by the red dot on the left and 
the red segment on the
  right. Here the system takes advantage of a shallow valley ($\mu$ is small)
  which provides an opportunity to accommodate the loading through low energy
  elastic deformation (remotely reminiscent of a fluid flow). Note, however,
  that the emergence of different types of instability mechanisms along direct
  and return deformation paths cannot be attributed solely to geometric
  nonlinearity of the Hadamard model, as a somewhat similar asymmetry is also
  present in the geometrically linear model whenever the two phases have
  different well ordered elastic moduli: in that case the material with larger
  elastic moduli plays the role of the high energy phase \cite{ak,gra,chbh07,achf16}.

In this paper our main technical approach is to generate bounds on the binodal
surface.  The simplest   bounds are obtained as a result of  probing the binodal by means of nucleating first rank laminates. Their optimality is proved by establishing their polyconvexity (and therefore quasiconvexity); the corresponding problem is algebraic  because the supporting null-Lagrangians can be constructed explicitly \cite{grtrpcx}. In contrast with the ``strongly solid''  regime of large $\mu$,  in the ``near-liquid''  regime of small $\mu$, not all of the first rank laminate bounds turn out to be  optimal. 

The  simple laminate bounds are first  improved by considering the nucleation of  second rank laminates even though,  as will be shown in
\cite{grtrliquid}, the second rank laminate bounds are also not optimal.
We could   improve them analytically  for hydrostatic strains only by considering 
nucleation of  a bounded circular inclusion in the infinite plane. Moreover, we provided a rationale behind the 
conjecture that the ``circular inclusion bound''  is in fact optimal. We show that if this conjecture could be proved, the 
the values of the deformation gradient in the exterior of the circular
nucleus would provide a bound on the whole binodal from the outside of the
binodal region. Another consequence of the conjectured optimality of the
inclusion-based nucleation bound  would be  the explicit formula for the quasiconvex envelope $QW(\BF)$ at all hydrostatic strains. To corroborate our conjecture we  juxtaposed the results obtained for 
 bounded
inclusions and unbounded second rank laminates and derived tight two-sided
bounds on the binodal.  As will be reported elsewhere, numerical computations
show that both bounds remain tight in the full range
of parameters for which they   are meaningful, with the hypothetical
bound being in perfect agreement with the numerically computed rank-one
convex binodal.

The paper is organized as follows. In Section~\ref{sec:prelim} we recapitulate   the 
general results from the calculus of variations for nonconvex vectorial
problems which are  later used in the  paper. In Section~\ref{sec:hadamat} we
specialize these results for the Hadamard material and present the numerical
illustrations of the obtained bounds. Analytical results for the limiting case
$\mu \to 0$ are presented in Section~\ref{sec:limcase} where we also compare
them with numerical computations. In Section~\ref{sec:hyp} we demonstrate the
far reaching consequences of the assumed optimality of the nucleation
bound. The paper ends with a general discussion and conclusions in
Section~\ref{sec:conc}.

\section{Preliminaries}
\setcounter{equation}{0}
\label{sec:prelim}
\emph{ Binodal region.} We recall that hyperelastic materials in an $n$-dimensional space are characterized by the following form of the energy stored in the deformed elastic body
\[
  E[\By]=\int_{\GO}W(\Grad\By(\Bx))d\Bx.
\]
Here $\GO\subset\bb{R}^{n}$ is the reference configuration, and $\By:\GO\to\bb{R}^{n}$ is the deformation. To deal with stable (i.e. experimentally observable) configurations of
the body one can replace the energy density $W(\BF)$ with a
relaxed one $QW(\BF)$, known as a quasiconvexification of $W(\BF)$. Even though, there is a formula for $QW(\BF)$ \cite{daco82}:
\begin{equation}
  \label{qcx}
  QW(\BF)=\inf_{\BGf\in C_{0}^{\infty}(B;\bb{R}^{n})}\nth{|D|}\int_{B}W(\BF+\Grad\BGf)d\Bx,
\end{equation}
 where $B$ is the unit ball, there are  no systematic approaches to actually compute it. A simpler, but just as useful
an object, is the elastic binodal.
\begin{definition}
\label{def:binodal}
  An elastic binodal is the boundary of the binodal region
  \begin{equation}
    \label{bindef}
        \mathfrak{B}=\{\BF:QW(\BF)<W(\BF)\}.
  \end{equation}
\end{definition}
\begin{definition}
  \label{def:stabF}
We say that the matrix $\BF$ is stable, if $W(\BF)=QW(\BF)$. 
\end{definition}
Thus, the binodal is the boundary separating the binodal region from the set
of stable points.

\emph{ Jump set. }While  there could be rank-one convex, non quasiconvex
functions, most cases of practical interest in elastic phase transitions
feature multiwell energies that are not rank-one convex. Such functions  possess a
non-trivial \emph{jump set}, stable points of which form a part of the binodal
(or the entire binodal, if one is lucky). The jump set is the set of solutions
$\BF=\BF_{-}$ of the equations \cite{grtrpe}
\begin{equation}
  \label{jsgen}
  \begin{cases}
      \BF_{+}=\BF_{-}+\Ba\otimes\Bn,\\
\jump{\BP}\Bn=0,\\
\jump{\BP^{T}}\Ba=0,\\
\jump{W}-\av{\lump{\BP},\jump{\BF}}=0,
  \end{cases}
\end{equation}
where $\Ba\not=0$,  $|\Bn|=1$, and 
the following standard notations are used.
\[
\BP_{\pm}=W_{\BF}(\BF_{\pm}),\quad\jump{\BF}=\BF_{+}-\BF_{-},\quad
\lump{\BP}=\frac{\BP_{+}+\BP_{-}}{2},\quad\av{\BA,\BB}=\Trc(\BA\BB^{T}),
\]
where $W_{\BF}$ indicates the matrix of partial derivatives $P_{ij}=\Md W/\Md
F_{ij}$. The vectors  $\Ba$ and $\Bn$ can be eliminated from \eqref{jsgen},
leaving a single scalar equation for $\BF$ that describes the jump set. 
The points on the jump set belong either to the binodal or to the binodal
region $\mathfrak{B}$, see \cite{grtrpe} for details. Hence, the jump set always represents a bound on the binodal region from within. 

One of the easy ways to detect the
unstable parts of the jump set is to use the Weierstrass condition, which is
necessary for stability.
\begin{equation}
  \label{WC}
  W^{\circ}(\BF,\Bb\otimes\Bm)\ge 0,\quad\forall\Bb\in\bb{R}^{n},\ |\Bm|=1,
\end{equation}
where
\[
W^{\circ}(\BF,\BH)=W(\BF+\BH)-W(\BF)-\av{W_{\BF}(\BF),\BH}.
\]
We have proved in \cite{grtrnc} that the pairs of points $\BF_{\pm}$ on the
jump set are either both stable or both unstable. Hence, a point $\BF_{+}$
satisfying (\ref{WC}) can be still classified as unstable, if $\BF_{-}$ fails
(\ref{WC}). While there are other conditions of stability that don't follow
from (\ref{WC}) (see \cite{grtrlhqcx}) we will only make use of an easily
verifiable corollary of (\ref{WC}) that restricts the rank-one test fields
$\Bb\otimes\Bm$ to an infinitesimally small \nbh\ of
$\jump{\BF}=\Ba\otimes\Bn$ (see \cite[(4.5)]{grtrpe}).

Currently, the only general tool for establishing stability of  an affine configuration $\BF$  is by proving polyconvexity of $W$ at $\BF$,
which is sufficient but rather far from necessary. In two dimensions it reduces
to finding a constant $m\in\bb{R}$, such that
\begin{equation}
  \label{pcxmethod}
  W^{\circ}(\BF,\BH)-m\det\BH\ge 0,\qquad\forall\BH\in\bb{R}^{2\times 2}.
\end{equation}
If (\ref{pcxmethod}) holds, then $\BF$ is stable in the sense of
Definition~\ref{def:stabF}. As shown in \cite{grtrpcx}, for points $\BF_{\pm}$ on the jump set  the only value of $m$ that could possibly work is, 
\begin{equation}
  \label{mjs}
  m=\frac{\av{\jump{\BP},\cof\jump{\BF}}}{{|\jump{\BF}|^{2}}}.
\end{equation}

\emph{Secondary jump set.}  The jump set described by
  (\ref{jsgen}) identifies the points on the binodal corresponding to
  nucleation of a layer of the new phase in the infinite domain occupied by the
original phase. As we have already mentioned, the entire jump set lies in the closure of
the binodal region \cite{grtrpe} and as such represents a bound on the binodal
from the inside. Another such bound is provided by testing stability of the
homogeneous phase with respect to nucleation of a twinned layer, or a second
rank laminate. Mathematically, we can treat it as a jump set of a partially
relaxed energy $\bra{W}$, defined by $\bra{W}(\bra{\BF})=\Gl
W(\BF_{+})+(1-\Gl)W(\BF_{-})$, where $\bra{\BF}=\Gl\BF_{+}+(1-\Gl)\BF_{-}$,
and where $\BF_{\pm}$ is the corresponding pair
on the jump set.
Thus, the secondary
jump set is defined by the system of equations
\begin{equation}
  \label{secjs}
  \begin{cases}
  \BF=\bra{\BF}+\Bb\otimes\Bm,\\
  \BP\Bm=\bra{\BP}\Bm,\\
\BP^{T}\Bb=\bra{\BP}^{T}\Bb,\\
W(\BF)-\bra{W}=\BP\Bm\cdot\Bb,
\end{cases}
\end{equation}
where 
\begin{equation}
  \label{bars}
  \bra{W}=\Gl W(\BF_{+})+(1-\Gl)W(\BF_{-}),\qquad
\bra{\BP}=\Gl\BP_{+}+(1-\Gl)\BP_{-},
\end{equation}
for some $\Gl\in[0,1]$, which also plays the role of a variable to be solved for in
(\ref{secjs}), along with $\BF$, $\Bb\not=0$, and $|\Bm|=1$.
It is clear that the so defined secondary
jump set represents another bound on the binodal region from within. 

\emph{Nucleation criterion.} Yet another method of probing the
binodal is to study  the nucleation of bounded inclusions either of a prescribed shape
\cite{blar,lba,kufr88} or of an optimal  shape which must be
determined \cite{karo72,pineau,kh}. The theory justifying why these tests
probe the binodal was developed in \cite{grtrmms}. In the case of ``nucleation
of a bounded inclusion'', the criterion for $\BF_{0}$ to be ``marginally
stable'', i.e. to lie in the closure of $\mathfrak{B}$, is the existence of a
field
\[
\BGf\in\CS=\{\BGf\in L^{2}_{\rm loc}(\bb{R}^{n}):\Grad\BGf\in L^{2}(\bb{R}^{n};\bb{R}^{n})\},
\]
such that
\begin{equation}
  \label{nucpde}
  \Div \BP(\BF_{0}+\Grad\BGf)=0,\qquad\Div\BP^{*}(\BF_{0}+\Grad\BGf)=0
\end{equation}
in the sense of distribution in $\bb{R}^{n}$, where
\[
\BP(\BF)=W_{\BF}(\BF),\qquad\BP^{*}(\BF)=W(\BF)\BI_{n}-\BF^{T}\BP(\BF),
\]
and where the solution $\BGf$ satisfies the non-degeneracy condition
\begin{equation}
  \label{ndgmms}
  \int_{\bb{R}^{n}}W_{\BF}^{\circ}(\BF_{0},\Grad\BGf)d\Bx\not=0.
\end{equation}
In the case of nucleation of a bounded inclusion $\Go$ with smooth boundary
the verification of (\ref{nucpde}) consists in checking that the field
$\BGf\in\CS$ solves $\Div\BP(\BF_{0}+\Grad\BGf)=0$ both inside and outside of
$\Go$, together with the condition that the traces
$\BF_{\pm}(\Bx)=\BF_{0}+\Grad\BGf_{\pm}(\Bx)$ on the two sides of $\Md\Go$
form a corresponding pair on the jump set for each $\Bx\in\Md\Go$. If, in
addition, we can somehow prove that $\BF+\Grad\BGf(\Bx)$ is stable in the sense
of Definition~\ref{def:stabF}, for each $\Bx\in\bb{R}^{n}$, then $\BF_{0}$
must lie on the binodal. Conversely, if it is known that $\BF_{0}$ is stable,
then all matrices $\BF(\Bx)=\BF_{0}+\Grad\BGf(\Bx)$ are stable for all
$\Bx\in\bb{R}^{n}$.

\section{The Hadamard material} 
\setcounter{equation}{0}
\label{sec:hadamat}
In this paper we focus   on   a particularly simple, yet  
nontrivial energy
\begin{equation}
  \label{enerex}
  W(\BF)=\frac{\mu}{2}|\BF|^{2}+h(d),\quad\BF\in\{\BF\in GL(n):\det\BF>0\},\quad d=\det\BF,
\end{equation}
where $h(d)$ is a $C^{2}(0,+\infty)$ function with a double-well shape. In our explicit computations and
illustrations we use the quartic double-well energy\footnote{Formula
  (\ref{quartich}) only needs to hold in an arbitrary \nbh\ of
  $[d_{1},d_{2}]$. The potential $h(d)$ can be modified outside of that \nbh\
  arbitrarily, as long as $h^{**}(d)=h(d)$ there. In particular, the
  singular behavior of $h(d)$ as $d\to 0^{+}$, required in nonlinear
  elasticity, can be easily assured.}
\begin{equation}
  \label{quartich}
  h(d)=(d-d_{1})^{2}(d-d_{2})^{2},
\end{equation}
which affords certain simplification of general formulas. In this section we  provide an approximation for the binodal of this
energy, even though its quasiconvex envelope is not known. We begin with the
computation of the jump set for this class of materials. 

\emph{The jump set.}  In Appendix~\ref{sec:genodgen} we summarize, for the
sake of completeness, the discussion of the jump set from 
\cite{grtrsolid},  which is adapted   to our two-dimensional setting.  

The main result of the Appendix~\ref{sec:genodgen} is that the jump set of
(\ref{enerex}) consists of matrices $\BF_{\pm}$, whose two singular values
labelled\footnote{We use the notation $\pm$ to make two statements at the same
  time, one for the ``$+$'' sign, the other for the ``$-$'' sign. For example,
  our statement says that the matrix $\BF_{+}$ has two
  singular values $\Gve_{0}$ and $\Gve_{+}$ and the matrix $\BF_{-}$ has two
  singular values $\Gve_{0}$ and $\Gve_{-}$.}  $\Gve_{0}$ and $\Gve_{\pm}$
satisfy the equations
\begin{equation}
  \label{js2D}
    \Gve_{0}\jump{h'}+\mu\jump{\Gve}=0,\quad
    \jump{h}-\lump{h'}\jump{d}=0,\quad d_{\pm}=\det\BF_{\pm}=\Gve_{0}\Gve_{\pm}.
  \end{equation}
 The first equation in (\ref{js2D}) is equivalent to (\ref{jumpS}) if
we recall the definition of $d_{\pm}$, given in the third equation in (\ref{js2D}).
Our notation reflects the fact that for each pair $\BF_{\pm}$ on the jump set there is a frame in which both matrices are diagonal and share the same singular value
$\Gve_{0}$ with the same eigenvector. 

Equations (\ref{js2D}) can be now used
to derive the semi-explicit parametric equations of the jump set,
with $d_{+}=\Gve_{0}\Gve_{+}$,  serving as a parameter. Given $d_{+}$
we can use the second equation in (\ref{js2D}) to
 solve for $d_{-}$. This solution will be  denoted $d_{-}=D(d_{+})$.
Multiplying the first equation in (\ref{js2D}) by $\Gve_{0}$ we obtain the parametric equations
\[
\begin{cases}
    \Gve_{0}(d_{+})=\sqrt{-\frac{\mu\jump{d}}{\jump{h'}}},\\
\Gve_{+}(d_{+})=\frac{d_{+}}{\Gve_{0}(d_{+})}.
\end{cases}
\]
 In the special case of potential \eqref{quartich} further simplifications can
 be made. For example,
\[
  \jump{h}-\lump{h'}\jump{d}=\jump{d}^{3}(d_{1}+d_{2}-d_{+}-d_{-}),
\]
and therefore, $d_{-}=d_{1}+d_{2}-d_{+}=:D(d_{+})$. It then follows that
\begin{equation}
  \label{quartsimp}
  \lump{h'}=0,\qquad\Gve_{+}+\Gve_{-}=\frac{d_{1}+d_{2}}{\Gve_{0}}.
\end{equation}
In particular, we can eliminate $h'(d_{\pm})$ using 
(\ref{js2D}) and (\ref{quartsimp}) to obtain:
\begin{equation}
  \label{hprimesmu}
  h'(d_{\pm})=\lump{h'}\pm\hf\jump{h'}=\mp\frac{\mu}{2}\frac{\jump{\Gve}}{\Gve_{0}}.
\end{equation}
Moreover, for quartic energy (\ref{quartich}) we can  write the equation of the jump
set explicitly as $\Gve_{\pm}=\Gve_{\pm}(\Gve_{0})$ where 
$\Gve_{\pm}=d_{\pm}/\Gve_{0}$, and $d_{\pm}$ solves
\begin{equation}
  \label{jsquart}
  (d_{\pm}-d_{1})(d_{\pm}-d_{2})=-\frac{\mu}{4\Gve_{0}^{2}}.
\end{equation}
The two roots of (\ref{jsquart}) are the values of $d_{\pm}$, where, by
convention, we denote by $d_{+}$ the larger root. Equation (\ref{jsquart}) has
exactly two real roots whenever $\Gve_{0}>\sqrt{\mu}/(d_{2}-d_{1})$.  Hence,
 explicitly,
\begin{equation}
  \label{jsexpmu}
  \Gve_{\pm}=\nth{2\Gve_{0}}\left(d_{1}+d_{2}\pm\sqrt{(d_{2}-d_{1})^{2}-\frac{\mu}{\Gve_{0}^{2}}}\right),\qquad\Gve_{0}\ge\frac{\sqrt{\mu}}{d_{2}-d_{1}}.
\end{equation}
In our calculations we will use equations (\ref{hprimesmu}) to eliminate all
occurrences of $h'(d_{\pm})$ and equations (\ref{jsexpmu}) to eliminate
$\Gve_{\pm}$, both uniquely determined by a single
parameter $\Gve_{0}$.
\begin{figure}[h!]
  \centering
\includegraphics[scale=0.35]{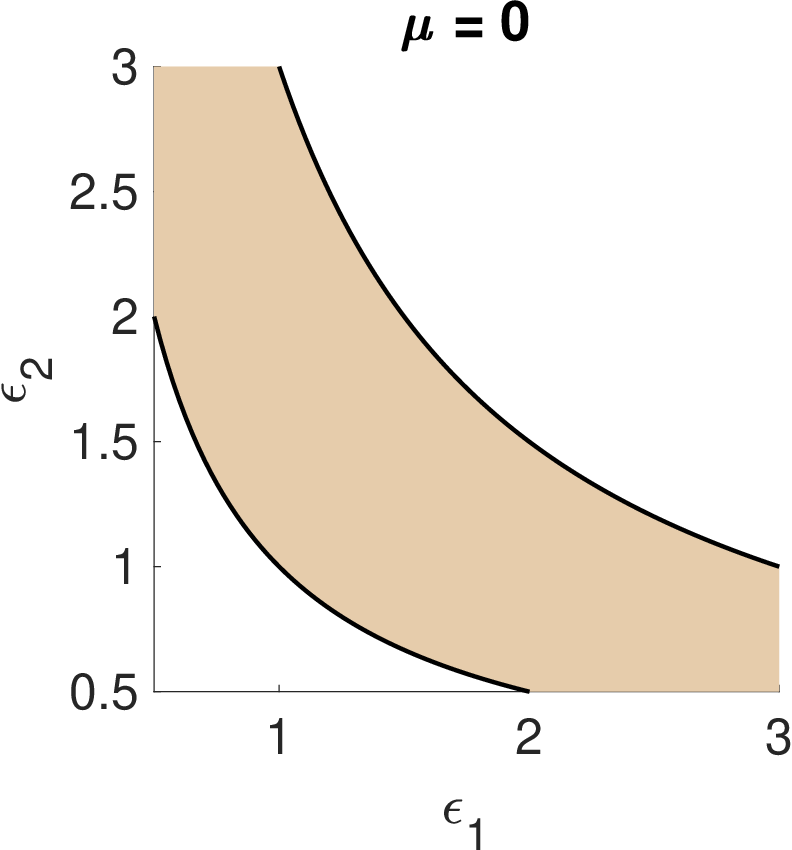}
  \includegraphics[scale=0.35]{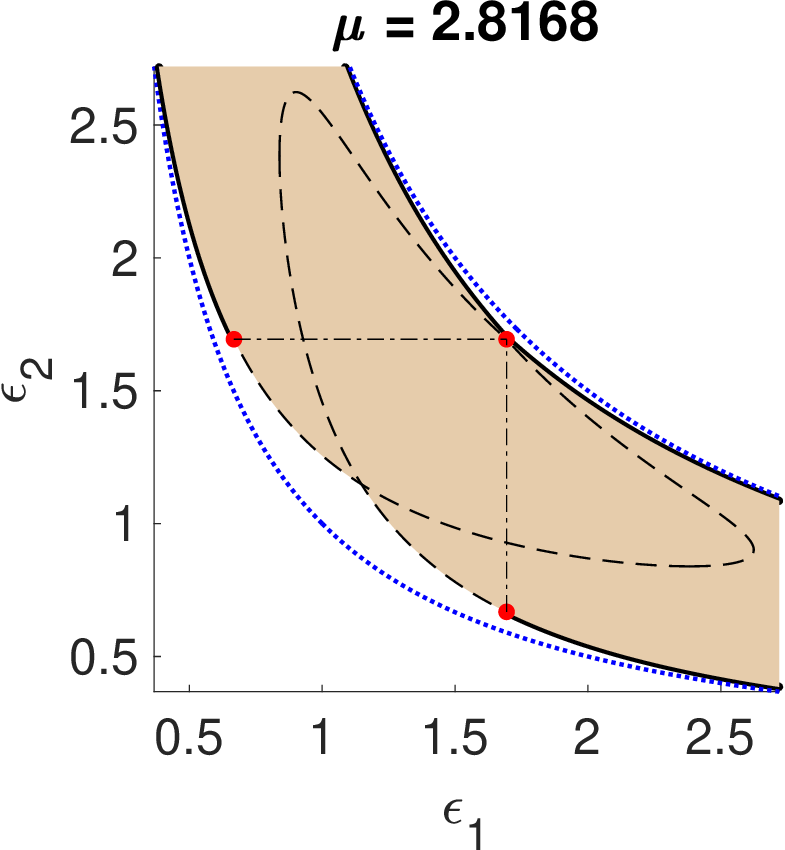}~~~~~
\includegraphics[scale=0.35]{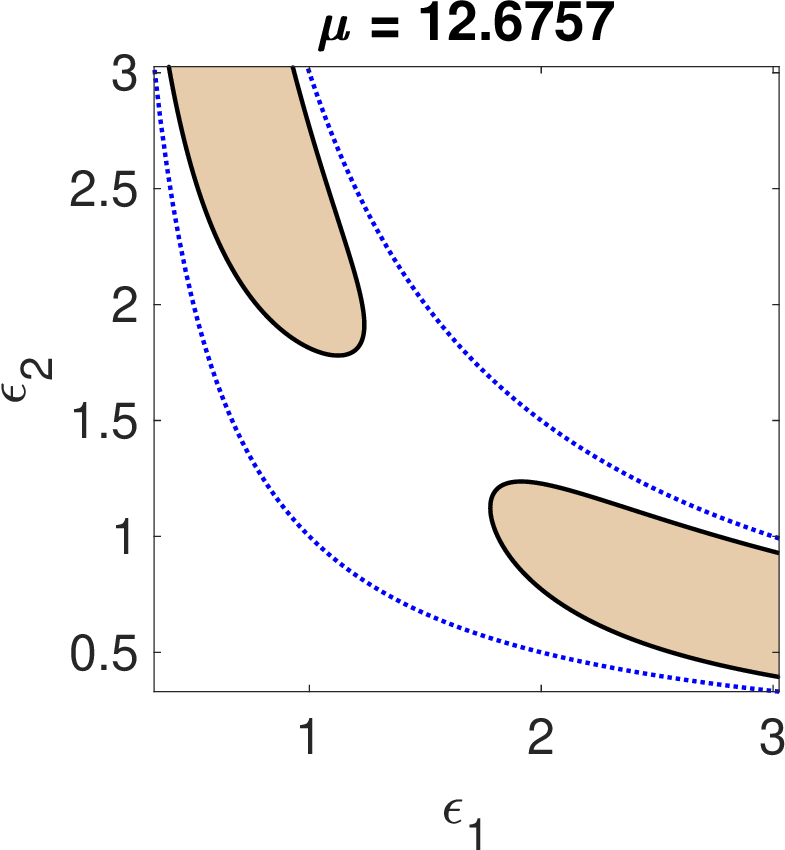}
\caption{Jump sets in the case when $h(d)$ is given by (\ref{quartich}) with
  $d_{1}=1$, $d_{2}=3$, and different values of $\mu$.  The convexification
  hyperbolas $\Gve_{1}\Gve_{2}=d_{1,2}$ are shown by blue dotted lines. The
  bold black lines show   stable part of the jump set. The
  dashed black lines represent the  unstable part of the jump set. The
  W-points are represented by red dots and dashed-dotted lines show the
  rank-one connections between the W-points. The shaded region is the part of
  the binodal region delimited by the jump set. Its interior is unstable.}
  \label{fig:jsets}
\end{figure}

\emph{Numerical illustrations.}  When $\mu$ is sufficiently large the jump set
is known to comprise the entire binodal \cite{grtrsolid}.  Since each point of
such binodal corresponds to the nucleation of a simple laminate, one
immediately obtains an explicit formula for the relaxation $QW(\BF)$. However,
as the shear modulus $\mu$ decreases, parts of the jump set may become
unstable which will also affect the structure of $QW(\BF)$. To illustrate this
point we show in Fig.~\ref{fig:jsets} the jump sets in the case when $h(d)$ is
given by (\ref{quartich}), and the three different values of the shear modulus
$\mu$ are chosen to be of the form $\mu=0$, $\mu_{\rm
  top}/3$,
and $1.5\mu_{\rm top}$, where $\mu_{\rm top}$ is the largest value of $\mu$
for which the jump set has points of self-intersection.  In
Fig.~\ref{fig:jsets} the dotted lines indicate ``convexification hyperbolas'',
i.e., hyperbolas $\Gve_{2}=d_{1}/\Gve_{1}$ and $\Gve_{2}=d_{2}/\Gve_{1}$,
where the interval $[d_{1},d_{2}]$ is the interval on which $h(d)$ differs
from its convex hull. The shaded region delimited by the jump set is unstable,
while all points outside of the region bounded by the convexification
hyperbolas are stable. It will be our main goal to specify the  precise
boundary of the unstable domain in the limit $\mu\to 0$.

\emph{W-points.}  In \cite{grtrlhqcx} we have shown that the easily computable corollary of the
Weierstrass condition (\ref{WC}) for the energy (\ref{enerex}) has the form
\begin{equation}
  \label{LHlam}
  \Gve_{0}\ge\Gve_{\pm}.
\end{equation}
In \cite{grtrsolid} we have shown that this condition is always satisfied for
large values of $\mu$ as is evident from the right panel in
Fig.~\ref{fig:jsets}, while it has an obvious geometric interpretation in the
middle panel in which the part of the jump set failing (\ref{LHlam}) is shown as
a dashed line. The points marked by red dots in Fig.~\ref{fig:jsets} that
delimit the part of the jump set satisfying (\ref{LHlam}) will be called the
Weierstrass points or W-points, for short. It will be shown in \cite{grtrliquid}
that the solid portion of the jump set delimited by W-points is polyconvex for
all sufficiently small $\mu$.

\emph{Polyconvexity of W-points.}
By their nature W-points are either unstable or delimit the boundary of
stability of the jump set. Our intuition, to be justified in
\cite{grtrliquid}, is that the larger the shear component of a point on the
jump set the more stable it is. Thus, the range of $\mu$ for which
W-points are polyconvex is also the range of $\mu$ for which the part of the
jump set with larger shear, delimited by W-points is polyconvex. 

One can  provide an almost explicit
characterization of all values of $\mu$ for which W-points  are also points of polyconvexity assuming the
quartic nonlinearity (\ref{quartich}). Indeed, as discussed above, in order to prove the polyconvexity of
W-points we need to establish (\ref{pcxmethod}), where $m$ is given by
(\ref{mjs}). This problem has been already analyzed in
\cite{grtrsolid},  and we briefly summarize here  the obtained 
  results  for the sake of completeness.

Establishing  (\ref{pcxmethod}) for the energy (\ref{enerex}) is
equivalent to showing that
\begin{equation}
  \label{Psifun}
\Psi(\BF)=\frac{\mu}{2}|\BF-\BF_{\pm}|^{2}+h(\det\BF)-h_{\pm}-
h'_{\pm}\av{\cof\BF_{\pm},\BF-\BF_{\pm}}-m\det(\BF-\BF_{\pm})
\end{equation}
is globally minimized by $\BF_{\pm}$. Our notation in (\ref{Psifun}) emphasizes the fact
that either choice of sign in $\BF_{\pm}$ results in one and the same function
$\Psi(\BF)$. 

We first observed that the minimizer of $\Psi(\BF)$ must be
a critical point, since $\Psi(\BF)\to+\infty$, when
$|\BF|\to\infty$. We then showed that at all points on the jump
set, except the points of self-intersection, the critical points $\BF$
must be diagonal in the same frame as $\BF_{\pm}$. Denoting by $x$ and
$y$ the two diagonal entries of $\BF$ we obtain
\[
\Psi(\BF)=\Phi(x,y)+{\rm const},
\]
where
\[
\Phi(x,y)=\frac{\mu}{2}(x^{2}+y^{2})-\Ga x-\Gb y+h(xy)-mxy,
\]
\begin{equation}
  \label{abR}
  \Ga=\mu(\Gve_{+}+\Gve_{-}),\quad
\Gb=\mu\left(\Gve_{0}+\frac{\Gve_{+}\Gve_{-}}{\Gve_{0}}\right),\quad m=\frac{\jump{h'd}}{\jump{d}}.
\end{equation}
When we minimized  $\Phi(x,y)$ over all $(x,y)$, such that $xy=d$ we  
concluded that the minimizer is $(d/y,y)$, where 
$y=y(d)$ is the largest root of 
\begin{equation}
   \label{xyquart0}
y^{4}-\Gb_{0} y^{3}+d\Ga_{0} y-d^{2}=0,\quad\Ga_{0}=\Gve_{+}+\Gve_{-},\quad
\Gb_{0}=\Gve_{0}+\frac{\Gve_{+}\Gve_{-}}{\Gve_{0}}.
 \end{equation}
Thus, the minimum of $\Phi(x,y)$ is achieved at a finite point corresponding
to a critical point of $\phi(d)=\Phi(d/y(d),y(d))$.

In the special case of W-points we have\footnote{ Technically, at the W-points
  there could be other, not necessarily diagonal critical states, however, by continuity, the diagonal critical points would still
deliver the global minimum of $\Psi(\BF)$.} $\Gve_{+}=\Gve_{0}$ and therefore
$\Ga_{0}=\Gb_{0}=\Gve_{-}+\Gve_{0}$. In this case equation (\ref{xyquart0})
factors
\begin{equation}
  \label{Wyeq}
  (y^{2}-d)(y^{2}-\Ga_{0}y+d)=0.
\end{equation}
The largest root is $y=\hf(\Ga_{0}+\sqrt{\Ga_{0}^{2}-4d})$, provided $0<d\le\Ga_{0}^{2}/4$.
If $d>\Ga_{0}^{2}/4$, then the quartic has only two real roots
$y=\pm\sqrt{d}$. Thus,
\[
y(d)=
\begin{cases}
(\Ga_{0}+\sqrt{\Ga_{0}^{2}-4d})/2,&d\le\Ga_{0}^{2}/4,\\
  \sqrt{d},&d>\Ga_{0}^{2}/4.
\end{cases}
\]
 Now,
\[
\phi(d)=\min_{y\in\bb{R}}\{\Phi_{0}(d/y,y)+h(d)-md\},\quad
\Phi_{0}(x,y)=\frac{\mu}{2}(x^{2}+y^{2})-\Ga x-\Gb y
\]
Therefore,
\[
\phi'(d)=\nth{y(d)}\dif{\Phi_{0}}{x}(d/y,y)+h'(d)-m=\mu\frac{d-\Ga_{0} y(d)}{y(d)^{2}}+h'(d)-m.
\]
In the case of W-points  for which $y(d)$ solves (\ref{Wyeq}) we see that 
\[
\frac{d-\Ga_{0} y(d)}{y(d)^{2}}=-1
\]
when $d\le\Ga_{0}^{2}/4$. Hence, any critical point of $\phi(d)$ in this
regime would have to satisfy
\[
h'(d)-\mu-m=0.
\]
One of the solutions is $d_{-}$, which always satisfies
$d_{-}\le\Ga_{0}^{2}/4$. If this equation has 3 solutions, the middle one
corresponds to a local maximum of $\phi(d)$, while the third $d^{*}>d_{+}$
always fails to satisfy $d^{*}\le\Ga_{0}^{2}/4$ because
$d_{+}=\Gve_{0}^{2}>(\Gve_{-}+\Gve_{0})^{2}/4$, due to (\ref{LHlam}). We
conclude that the only critical points of $\phi(d)$ that need to be checked
are the ones that satisfy $d>\Ga_{0}^{2}/4$.  In this regime $y(d)^{2}=d$, and
\[
\phi'(d)=\mu\left(1-\frac{\Ga_{0}}{\sqrt{d}}\right)+h'(d)-m.
\]
Observe that $\phi'(d)>0$
when $d\ge\max(\Ga_{0}^{2},\Tld{d}_{+})$, where $\Tld{d}_{+}$ is the largest
root of $h'(d)-m$. Hence we only need to check for critical points in a
specific bounded interval. In fact, if $h(d)$ is given by (\ref{quartich}),
then it is easy to see that $\phi'(d)>0$ for all $d\ge\Ga_{0}^{2}$. Hence, we
only need to check for critical points of $\phi(d)$ on
$(\Ga_{0}^{2}/4,\Ga_{0}^{2})$. In addition, since $\lump{h'}=0$, for $h(d)$
given by (\ref{quartich}), we have
$m=-\mu\lump{\Gve}/\Gve_{0}=-\mu\Ga_{0}/(2\Gve_{0})$. Thus, we obtain
the following characterization of polyconvexity of W-points.
\begin{theorem}
  \label{th:pcxW}
Let $h(d)$ be given by (\ref{quartich}), then W-points are polyconvex
whenever
\begin{equation}
  \label{Wpcxcrit}
\min_{d\in\left[\frac{\Ga_{0}^{2}}{4},\Ga_{0}^{2}\right]}\left(h(d)+\mu\left(d+\frac{\Ga_{0}d}{2\Gve_{0}}
-2\Ga_{0}\sqrt{d}\right)\right)=h(\Gve_{0}^{2})
-\mu\Gve_{0}\left(\frac{\Gve_{0}}{2}+\frac{3\Gve_{-}}{2}\right).
\end{equation}
where $\Ga_{0}=\Gve_{0}+\Gve_{-}$, with $(\Gve_{0},\Gve_{-})$,
$(\Gve_{-},\Gve_{0})$, and $(\Gve_{0},\Gve_{0})$ being the coordinates of W-points.
\end{theorem}
The \rhs\ in (\ref{Wpcxcrit}) is just $\phi(\Gve_{0}^{2})$, where $\phi(d)$ is
the function being minimized in (\ref{Wpcxcrit}). For quartic energy
(\ref{quartich}) we compute the coordinates of W-points by solving
\[
  -4d(d-d_{1})(d-d_{2})=\mu.
\]
Then $\Gve_{0}^{2}$ is the largest root, and 
\[
\Gve_{-}=\frac{d_{1}+d_{2}-\Gve_{0}^{2}}{\Gve_{0}}.
\]
We can compute the largest value of $\mu$ for which (\ref{Wpcxcrit}) holds by
substituting $\mu=-4\Gve_{0}^{2}(\Gve_{0}^{2}-d_{1})(\Gve_{0}^{2}-d_{2})$ into
(\ref{Wpcxcrit}) and regarding $\Gve_{0}\le\sqrt{d_{2}}$ as a parameter. When
$\Gve_{0}=\sqrt{d_{2}}$, $\phi(d)-\phi(d_{2})$ is a positive polynomial in
$x=\sqrt{d}$. We then seek numerically the largest value of
$\Gve_{0}<\sqrt{d_{2}}$ for which the polynomial
$P(x)=(\phi(x^{2})-\phi(\Gve_{0}^{2}))/(x-\Gve_{0})^{2}$ develops a double
root. Algebraically this means seeking the largest root
$\Gve_{0}<\sqrt{d_{2}}$ of the discriminant. This solution
gives the largest value of $\mu$ below which the W-points are polyconvex. For
example, when $d_{1}=1$, $d_{2}=3$, we have polyconvexity of W-points for all
$\mu<6.35888$. When $\mu$ increases past that value it enters a regime where
the W-points are no longer polyconvex, but could still be
quasiconvex. Increasing $\mu$ even further, we enter a regime where W-points
fail a more sophisticated stability test from \cite{grtrlhqcx}. The exact
value of $\mu$, where W-points stop being stable is unknown.
In this paper we are working exclusively in the regime of sufficiently small
$\mu$, when W-points are polyconvex.

\emph{Secondary jump set.}  For general values of $\mu$ the algebraic equations (\ref{secjs}) describing the secondary jump set can
  be solved only numerically. By contrast, when $\mu$ is small, the
asymptotics of the solutions can be obtained  explicitly, providing an
excellent approximation to the computed secondary jump set for $\mu<3$, with
$d_{1}=1$, $d_{2}=3$. While the entire secondary jump set is 
unstable,  as will be proved in \cite{grtrliquid}, we
will see that it provides an excellent (inside)  bound for the binodal. 
 Here we specialize general equations (\ref{secjs}) to the
  specific energy density (\ref{enerex}) without assuming that the rigidity
  measure $\mu$ is small.

Suppose that $\BF_{0}$ lies on the secondary jump set. Then there exists
$\Gve_{\pm}$, $y$ and $\Gl\in[0,1]$, such that the pair $\BF_{0},\bra{\BF}$, where
\[
\bra{\BF}=\mat{\bra{\Gve}}{0}{0}{\Gve_{0}},\qquad\bra{\Gve}=\Gl \Gve_{+}+(1-\Gl)\Gve_{-},
\]
satisfies the secondary jump set equations (\ref{secjs}).
We compute
\[
\bra{\BP}=\Gl\BP_{+}+(1-\Gl)\BP_{-}=\mat{\mu\bra{\Gve}+\bra{h'}\Gve_{0}}{0}{0}{\mu\Gve_{0}+\bra{\Gve h'}}.
\]
We have
\[
\BP_{0}=\mu\BF_{0}+h'(d_{0})\cof\BF_{0}=\mu\mat{\bra{\Gve}}{0}{0}{\Gve_{0}}+\mu\Bb\otimes\Bm+
h'(d_{0})\left(\mat{\Gve_{0}}{0}{0}{\bra{\Gve}}+\Bb^{\perp}\otimes\Bm^{\perp}\right).
\]
Thus, the second and the third equations in the secondary jump set system (\ref{secjs}) become
\[
\begin{cases}
  \mat{(h'(d_{0})-\bra{h'})\Gve_{0}}{0}{0}{h'(d_{0})\bra{\Gve}-\bra{\Gve h'}}\Bm=-\mu\Bb,\\[6ex]
\mat{(h'(d_{0})-\bra{h'})\Gve_{0}}{0}{0}{h'(d_{0})\bra{\Gve}-\bra{\Gve h'}}\Bb=-\mu|\Bb|^{2}\Bm.
\end{cases}
\]
These equations result in 3 possibilities:
\begin{enumerate}
\item[(a)] $(h'(d_{0})-\bra{h'})\Gve_{0}=h'(d_{0})\bra{\Gve}-\bra{\Gve h'}=-\Gg$, 
$\mu\Bb=\Gg\Bm$, $\Bm\in\bb{S}^{1}$;
\item[(b)] $(h'(d_{0})-\bra{h'})\Gve_{0}=-(h'(d_{0})\bra{\Gve}-\bra{\Gve h'})=-\Gg$,
  $\mu\Bb=\Gg\BI_{-}\Bm$,  $\BI_{-}=\mat{1}{0}{0}{-1}$, $\Bm\in\bb{S}^{1}$;
\item[(c)] $(h'(d_{0})-\bra{h'})\Gve_{0}\not=\pm(h'(d_{0})\bra{\Gve}-\bra{\Gve h'})$.
\end{enumerate}
Possibility $(c)$ implies that $\BF_{0}$ must be diagonal, and will be our
main focus. In \cite{grtrliquid}  it will be shown that in the cases  (a) and (b) there are no solutions.
Let us therefore assume that $\BF_{\pm}$ is diagonal and has the form
\[
\BF_{\pm}=\mat{\Gve_{\pm}}{0}{0}{\Gve_{0}}.
\]
This implies that $\bra{\BF}-\BF_{0}=\Gb\tns{\Be_{2}}$.
In particular
\[
\BF_{0}=\mat{x_{0}}{0}{0}{y_{0}},\quad x_{0}=\bra{\Gve}=\Gl\Gve_{+}+(1-\Gl)\Gve_{-},
\quad\Gl\in(0,1).
\]
Let us compute the diagonal matrices $\BP_{\pm}$ using equations
(\ref{quartsimp}) and (\ref{hprimesmu}).
\[
P_{\pm}^{11}=\mu\Gve_{\pm}+h'(d_{\pm})\Gve_{0}=\mu\lump{\Gve}=\frac{\mu(d_{1}+d_{2})}{2\Gve_{0}},\quad
P_{\pm}^{22}=\mu\Gve_{0}+h'(d_{\pm})\Gve_{\pm}=\mu\left(\Gve_{0}\mp\frac{\jump{\Gve}\Gve_{\pm}}{2\Gve_{0}}\right).
\]
Let us compute the diagonal matrix $\BP_{0}$.
\[
P_{0}^{11}=\mu x_{0}+h'(d_{0})y_{0}=\mu\bra{\Gve}+h'(d_{0})\frac{d_{0}}{\bra{\Gve}},\quad
P_{0}^{22}=\mu y_{0}+h'(d_{0})x_{0}=\frac{\mu d_{0}}{\bra{\Gve}}+h'(d_{0})\bra{\Gve}.
\]
Traction continuity equation $(\bra{\BP}-\BP_{0})\Be_{2}=0$ then becomes
\[
\Gve_{0}+\frac{\jump{\Gve}}{2\Gve_{0}}(\Gve_{-}-2\Gl\lump{\Gve})-\frac{d_{0}}{\bra{\Gve}}-\frac{h'(d_{0})}{\mu}\bra{\Gve}=0.
\]
It will be convenient to use $\bra{\Gve}$ as a variable in place of
$\Gl$. Replacing $\Gl$ above using $\bra{\Gve}=\Gve_{-}+\Gl\jump{\Gve}$ we
obtain
\begin{equation}
  \label{trac}
\frac{d_{0}}{\bra{\Gve}}=\Gve_{0}+\frac{1}{\Gve_{0}}(\Gve_{+}\Gve_{-}-\lump{\Gve}\bra{\Gve})
-\frac{h'(d_{0})}{\mu}\bra{\Gve}.
\end{equation}

Let us now compute all the terms in the last equation in (\ref{secjs}).
\[
W(\BF_{0})=\frac{\mu}{2}(\bra{\Gve}^{2}+y_{0}^{2})+h(d_{0})=
\frac{\mu}{2}\left(\bra{\Gve}^{2}+\frac{d_{0}^{2}}{\bra{\Gve}^{2}}\right)+h(d_{0}).
\]
Next we compute
\[
\bra{W}=W_{-}+\Gl\jump{W}=W_{-}+\Gl\mu\jump{\Gve}\lump{\Gve}=W_{-}+\mu(\bra{\Gve}-\Gve_{-})\lump{\Gve},
\]
where $\jump{h}=-\lump{h'}\jump{d}=0$ has been used. 
We compute
\[
h(d_{-})=[(d_{-}-d_{1})(d_{-}-d_{1})]^{2}=\frac{\mu^{2}}{16\Gve_{0}^{4}},
\]
according to (\ref{jsquart}). Therefore,
\[
W_{-}=\frac{\mu}{2}(\Gve_{-}^{2}+\Gve_{0}^{2})+\frac{\mu^{2}}{16\Gve_{0}^{4}}.
\]
We then compute $\BF_{0}-\bra{\BF}=(y_{0}-\Gve_{0})\tns{\Be_{2}}$. Therefore
\[
\av{\bra{\BP},\BF_{0}-\bra{\BF}}=\mu\left(\frac{d_{0}}{\bra{\Gve}}-\Gve_{0}\right)
\left(\Gve_{0}+\frac{1}{\Gve_{0}}(\Gve_{+}\Gve_{-}-\lump{\Gve}\bra{\Gve})\right).
\]
Finally, the Maxwell equation
$W(\BF_{0})-\bra{W}=\av{\bra{\BP},\BF_{0}-\bra{\BF}}$ can be written as
\begin{equation}
  \label{Maxwell}
  \hf\left(\bra{\Gve}^{2}+\frac{d_{0}^{2}}{\bra{\Gve}^{2}}\right)+\frac{h(d_{0})}{\mu}-
(\bra{\Gve}-\Gve_{-})\lump{\Gve}-\hf(\Gve_{-}^{2}+\Gve_{0}^{2})-\frac{\mu}{16\Gve_{0}^{4}}=
\left(\frac{d_{0}}{\bra{\Gve}}-\Gve_{0}\right)
\left(\Gve_{0}+\frac{1}{\Gve_{0}}(\Gve_{+}\Gve_{-}-\lump{\Gve}\bra{\Gve})\right).
\end{equation}
Next, in (\ref{Maxwell}) we replace $d_{0}/\bra{\Gve}$ by its expression
from (\ref{trac}).  As a result of such substitution the Maxwell relation will
also become a quadratic equation in $\bra{\Gve}$. This permits us to eliminate
this variable as a rational expression in terms of $\Gve_{0}$ and $d_{0}$.
The Maxwell relation will then reduce to a rational relation between $d_{0}$
and $\Gve_{0}$.
\begin{figure}[t]
  \centering
  \includegraphics[scale=0.5]{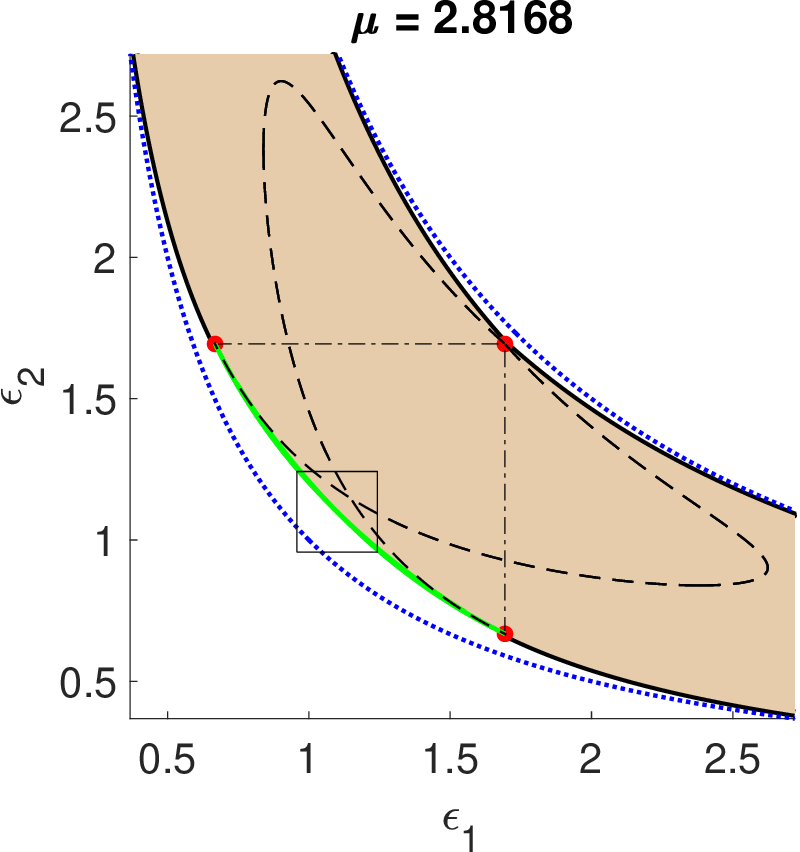}~~~~~~~
\includegraphics[scale=0.5]{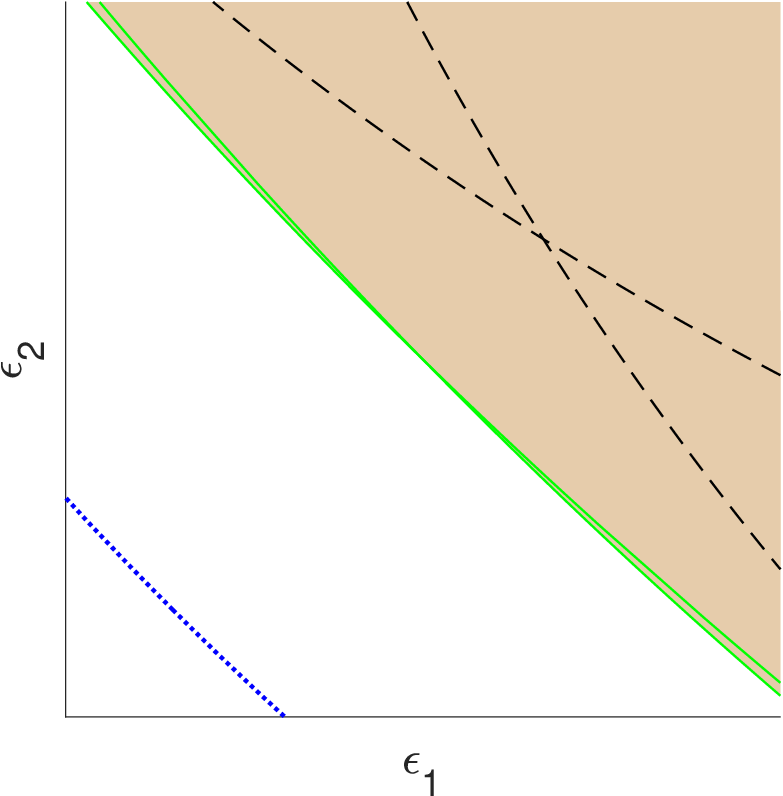}
   \caption{Secondary jump set (intersecting  green lines) computed   numerically from
     (\ref{trac}) and (\ref{Maxwell}). The right panel shows the blown-up
     central box in the left panel.}
  \label{fig:secjs}
\end{figure}

The implied  calculation can only be done with the aid of a computer algebra system,
since the remaining equation $F(\Gve_{0},d_{0})=0$ is very long and
complicated. For a given choice of numerical values of $\mu$, $d_{1}$ and
$d_{2}$ we can then solve $F(\Gve_{0},d_{0})=0$ numerically and   extract the
solutions which satisfy $\Gl\in[0,1]$. The result for $d_{1}=1$, $d_{2}=2$ and
$\mu=\mu_{\rm top}/3$ is shown as a green curve in
Fig.~\ref{fig:secjs}. As we can see, it identifies all points between the green
curve and the dashed lines of the primary jump set as a part of the binodal
region---an improvement over the primary jump set bound. 

As shown in the right panel of Fig.~\ref{fig:secjs}, the secondary jump set
consists of two curves related by symmetry with respect to the bisector of the
first quadrant.  
Each curve starts at a W-point and ends at a point (not marked) on the dashed
part of the jump set. The endpoints of the secondary jump set correspond to
the extreme values 0 and 1 of the volume fraction $\Gl$ in (\ref{bars}) and
must lie on the primary jump set. There are two possibilities. Either
$\BF\not=\bra{\BF}$ or $\BF=\bra{\BF}$ at $\Gl=0$ or 1. In the former case the
limiting position $\BF_{+}$ of $\bra{\BF}$ is rank-one related to two
different points on the jump set: $\BF_{-}$ (layer normal $\Be_{1}$) and $\BF$
(layer normal $\Be_{2}$). The W-point $\BF_{+}$ is the only one with this
property. All other points $\BF_{+}$ on the jump set have a unique counterpart
$\BF_{-}$.  In the latter case a detailed asymptotic analysis shows that that
the common limit point of $\BF$ and $\bra{\BF}$ must achieve equality in the
``Legendre-Hadamard for phase boundaries'' inequality from
\cite{grtrlhqcx}. When $\mu$ is small, this point lies on the dashed part of
the jump set and is used in numerical calculations. The technical details of
the analysis will be reported elsewhere. The sections of each of the two
curves from the W-point to the bisector of the quadrant form a part of the
boundary of the new shaded region of unstable points.

\emph{Circular nucleus.} As we have already mentioned, the secondary jump set
(shown in green in Fig.~\ref{fig:secjs}) is unstable \cite{grtrliquid}. That
means that the corresponding bound on the binodal is not optimal.
To improve this  bound we can use another method of probing
the binodal: nucleation of bounded equilibrium energy-neutral inclusions. The
theory justifying why such nucleation tests probe the binodal was developed in
\cite{grtrmms}. 

In the case of the isotropic, objective energy (\ref{enerex}) and a
hydrostatic loading it would be natural to assume that the shape of an optimal
precipitate is circular. The deformation gradient $\BF_{0}$ inside a circular
precipitate must be a constant hydrostatic field that jumps across the
circular boundary of the inclusion to fields $\BF(\Bx)$. In order for such a
configuration to be able to probe the binodal, $\BF_{0}$ and $\BF(\Bx)$ must
be corresponding pairs on the jump set. There is only one point on the jump
set satisfying these requirements $\BF_{0}=\Gve_{0}^{\rm W}\BI_{2}$, where
$(\Gve_{0}^{\rm W},\Gve_{0}^{\rm W})$ is the W-point that lies on the quadrant
bisector. As required, the field $\BF_{0}$ is rank-one connected to an
infinite family of fields
\[
\BF_{\BR}=\BR\mat{\Gve_{-}^{\rm W}}{0}{0}{\Gve_{0}^{\rm W}}\BR^{T},\qquad\BR\in SO(2),
\]
where $(\Gve_{-}^{\rm W},\Gve_{0}^{\rm W})$ is a coordinate of one of the other W-points. 
The deformation gradient outside of the circular inclusion must solve an
Euler-Lagrange equation for the energy (\ref{enerex}) 
\begin{equation}
  \label{liqpde}
\mu\GD\By+(\cof\Grad\By)\Grad h'(\det\Grad\By)=0,\qquad\Bx\in\bb{R}^{2}\setminus B(\Bzr,1),
\end{equation}
and agree with
$\BF_{\BR}$ at the point $\BR\Be_{1}$ on the boundary of the circular
inclusion:
\begin{equation}
  \label{sntbc}
  \Grad\By(\Bx)=\Gve_{-}^{\rm W}\tns{\Bn}+\Gve_{0}^{\rm
    W}\tns{\Bn^{\perp}},\qquad\Bx\in\Md B(\Bzr,1).
\end{equation}
Under these conditions,  both equations (\ref{nucpde}) will be satisfied for the possibly marginally stable
matrix
\[
\BF_{\infty}=\lim_{|\Bx|\to\infty}\Grad\By(\Bx)=\Gve_{\infty}\BI_{2}.
\] 
We also know that that the values of $\Grad\By(\Bx)$ inside the circular inclusion and its trace on the outside boundary of the inclusion must be  stable. Our results from \cite{grtrmms} then say that there are two possibilities. The first one is that   $\BF_{\infty}$ lies on the binodal and all values $\Grad\By(\Bx)$ in the exterior of the inclusion are stable. The second option is that  $\BF_{\infty}$ lies in the interior of the binodal region $\mathfrak{B}$. 

In our special radially symmetric case
we look for a radially symmetric solution of (\ref{liqpde})
\[
\By=\eta(r)\hat{\Bx},\qquad |\Bx|>1.
\]
The  unknown function $\eta(r)$ must solve
\begin{equation}
  \label{radequil}
  \begin{cases}
    \frac{\eta}{r}\frac{d}{dr}h'\left(\frac{\eta\eta'}{r}\right)+
\mu\left(\eta'+\frac{\eta}{r}\right)'=0,&r>1,\\
  \eta'(1)=\Gve_{-}^{\rm W},\qquad\eta(1)=\Gve_{0}^{\rm W}.
  \end{cases}
\end{equation}
The nonlinear second order ODE (\ref{radequil}) cannot be integrated
explicitly, but can be solved numerically. In order to do so, we need to
convert the infinite range $r>1$ into a finite one by means of the change of
the independent variable $x=1/r^{2}$. It will also be convenient to change the
dependent variable $v=\eta/r$, so that $v(x)$ would have a finite limit, when
$x\to 0$. Then $v(x)$ solves
\begin{equation}
  \label{nucodev}
  v''=-\frac{(v')^{2}vh''(v^{2}-2xvv')}{\mu+v^{2}h''(v^{2}-2xvv')},\quad x\in[0,1],\quad
v(1)=\Gve_{0}^{\rm W},\quad v'(1)=\frac{\Gve_{0}^{\rm W}-\Gve_{-}^{\rm W}}{2}.
\end{equation}
The value $\Gve_{\infty}=v(0)\BI_{2}$, which was found numerically, is shown as a blue dot in Fig.~\ref{fig:binodal1}. While we still cannot say whether the corresponding value $\BF_{\infty}$ indeed lies in the binodal, we obtained  an improved bound on the binodal compared to the secondary jump set (green line in Fig.~\ref{fig:binodal1}) by showing that hydrostatic strains between the blue dot and the green line are unstable. 
The conclusion holds, provided the non-degeneracy condition (\ref{ndgmms}) is verified.
A direct calculation shows that
\[
\int_{\bb{R}^{2}}W_{\BF}^{\circ}(\BF_{0},\Grad\BGf)d\Bx=-\BI_{2}\int_{\bb{R}^{2}}h''(\Gve_{\infty}^{2}) \Gve_{\infty}^{2}\left(\eta'(r)+\frac{\eta(r)}{r}-2\Gve_{\infty}\right)d\Bx.
\]
Thus,
\[
\int_{\bb{R}^{2}}W_{\BF}^{\circ}(\BF_{0},\Grad\BGf)d\Bx=-2\pi h''(\Gve_{\infty}^{2}) \Gve_{\infty}^{2}\BI_{2}\lim_{r\to\infty}(r\eta(r)-\Gve_{\infty}r^{2}).
\]
To see that the limit above exists and is non-zero, at least for sufficiently small $\mu>0$,
we simply solve (\ref{radequil}) for $\mu=0$, for which $\Gve_{0}^{\rm W}=\sqrt{d_{2}}$,
$\Gve_{-}^{\rm W}=\frac{d_{1}}{\sqrt{d_{2}}}$. The solution is
$\eta(r)=\sqrt{d_{1}r^{2}+d_{2}-d_{1}}$, and we easily see that
\[
\lim_{r\to\infty}(r\eta(r)-\Gve_{\infty}r^{2})=\frac{d_{2}-d_{1}}{2\sqrt{d_{1}}}.
\]
Hence, the non-degeneracy condition (\ref{ndgmms}) will hold for
sufficiently small $\mu>0$. The non-degeneracy will also hold for all $\mu$
below the topological transition, because if we write
$\Tld{\eta}(r)=\eta(r)-\Gve_{\infty}r$, then (assuming that $\Tld{\eta}'(r)\to
0$, as $r\to\infty$) $\Tld{\eta}(r)$ will solve, when $r$ is large, the
differential equation
\[
\Gve_{\infty}h''(\Gve_{\infty}^{2})\left(\Gve_{\infty}\left(\Tld{\eta}'+\frac{\Tld{\eta}}{r}\right)+\frac{\Tld{\eta}'\Tld{\eta}}{r}\right)+\mu\left(\Tld{\eta}'+\frac{\Tld{\eta}}{r}\right)=0.
\]
This integrates to
\[
\Gve_{\infty}h''(\Gve_{\infty}^{2})(2\Gve_{\infty}r\Tld{\eta}+\Tld{\eta}^{2})+2\mu r\Tld{\eta}=2c.
\]
Since $\Tld{\eta}$, satisfying $\Tld{\eta}'(r)\to 0$, as $r\to\infty$, cannot
be zero (it is the leading term of $\eta(r)-\Gve_{\infty}r$), we conclude that
the constant of integration $c$ cannot be zero either. Hence, we obtain that
\[
\lim_{r\to\infty}(r\eta(r)-\Gve_{\infty}r^{2})=\lim_{r\to\infty}r\Tld{\eta}(r)=\frac{c}{\mu+\Gve_{\infty}^{2}h''(\Gve_{\infty}^{2})}\not=0.
\]

\emph{Polyconvexity limits along $\Gve\BI_{2}$.}  In an attempt to
 prove stability of the point
$\Gve_{\infty}\BI_{2}$ we turn to the  problem of
  polyconvexity at points $\BF=\Gve\BI_{2}$. The problem reduces to finding 
 a constant $m\in\bb{R}$, such that (\ref{pcxmethod}) holds.
For our energy we compute
\[
W^{\circ}(\BF,\BH)=\frac{\mu}{2}|\BH|^{2}+h(\Gve^{2}+d+\Gve\Gth)-h(\Gve^{2})-\Gve
h'(\Gve^{2})\Gth,\qquad \Gth=\Trc\BH,\ d=\det\BH.
\]
We also have
\[
|\BH|^{2}=4s^{2}-2d+\Gth^{2},
\]
where
\[
\hf(\BH-\BH^{T})=\mat{0}{-s}{s}{0}.
\]
The set of all admissible values of $(\Gth,d,s)$ is described by the
inequality\footnote{This inequality is equivalent to $|\dev{\BH}|^{2}\ge0$,
  where $2\dev{\BH}=\BH+\BH^{T}-(\Trc\BH)\BI_{2}$.} $s^{2}\ge d-\Gth^{2}/4$.
Thus, proving that $W^{\circ}(\Gve\BI_{2},\BH)\ge m\det\BH$ for all
$\BH$ is equivalent to proving that
\[
2\mu\max\left\{0,d-\frac{\Gth^{2}}{4}\right\}+ 
\frac{\mu\Gth^{2}}{2}+h(\Gve^{2}+d+\Gve\Gth)-h(\Gve^{2})-\Gve
h'(\Gve^{2})\Gth\ge(m+\mu)d.
\]
Establishing this inequality splits into two cases
\begin{equation}
  \label{pcxineq}
  \frac{\mu\Gth^{2}}{2}+h(\Gve^{2}+d+\Gve\Gth)-h(\Gve^{2})-\Gve
h'(\Gve^{2})\Gth\ge (m+\mu)d,\quad\forall d\le\frac{\Gth^{2}}{4},
\end{equation}
and 
\begin{equation}
  \label{pcxine1}
  h(\Gve^{2}+d+\Gve\Gth)-h(\Gve^{2})-\Gve
h'(\Gve^{2})\Gth\ge (m-\mu)d,\quad\forall d\ge\frac{\Gth^{2}}{4}.
\end{equation}
In particular both inequalities must hold for $d=\Gth^{2}/4$. In that
case we must have
\begin{equation}
  \label{mstar}
  m\le\mu+4\min_{\Gth\in\bb{R}}\frac{h(\Gve^{2}+\Gth^{2}/4+\Gve\Gth)-h(\Gve^{2})-\Gve
h'(\Gve^{2})\Gth}{\Gth^{2}}=m^{*}.
\end{equation}
Changing variables $\Gd=\Gve^{2}+d+\Gve\Gth$ we obtain that $\BF=\Gve\BI_{2}$
is polyconvex \IFF there exists $m\le m_{*}$, such that
\begin{equation}
  \label{F1case}
  \inf_{\Gd\le(\Gve+\Gth/2)^{2}} F_{1}(\Gd,\Gth)\ge 0,\quad
F_{1}(\Gd,\Gth)=\frac{\mu\Gth^{2}}{2}+h(\Gd)-h(\Gve^{2})-\Gve
h'(\Gve^{2})\Gth-(m+\mu)(\Gd-\Gve\Gth-\Gve^{2}),
\end{equation}
and
 \begin{equation}
  \label{F2case}
  \inf_{\Gd\ge(\Gve+\Gth/2)^{2}} F_{2}(\Gd,\Gth)\ge 0,\quad
F_{2}(\Gd,\Gth)=h(\Gd)-h(\Gve^{2})-\Gve
h'(\Gve^{2})\Gth-(m-\mu)(\Gd-\Gve\Gth-\Gve^{2}).
\end{equation}
The case (\ref{F2case}) is clear, because $F_{2}(\Gd,\Gth)$ is linear in $\Gth$
and the minimum is always achieved on the boundary of the admissible domain,
i.e. $\Gd=(\Gve+\Gth/2)^{2}$ or, equivalently, $d=\Gth^{2}/4$. In this case
inequality (\ref{F2case}) holds whenever $m\le m^{*}$.

The function $F_{1}(\Gd,\Gth)$ is quadratic in $\Gth$ and therefore achieves
its minimal value either on the boundary, corresponding to $d=\Gth^{2}/4$ or
at the critical point, satisfying
\begin{equation}
  \label{critpt}
\Gth=\frac{\Gve(h'(\Gve^{2})-m-\mu)}{\mu},\quad h'(\Gd)=m+\mu,
\end{equation}
provided $\Gd\le (\Gth/2+\Gve)^{2}$, which holds \IFF
\begin{equation}
  \label{dconstr}
  \Gd\le\frac{\Gve^{2}(h'(\Gve^{2})+\mu-m)^{2}}{4\mu^{2}}.
\end{equation}
We remark that taking $\Gth=-4\Gve$ in (\ref{mstar}) we infer that
$m^{*}\le\mu+h'(\Gve^{2})$. Thus, the \rhs\ of (\ref{dconstr}) is monotone
decreasing in $m$, when $m\le m^{*}$.

Let $\Tld{d}_{1}<\Tld{d_{2}}$ be the two inflection points of $h(d)$. The
point $\Tld{d}_{1}$ is the point of local maximum of $h'(d)$, while
$\Tld{d}_{2}$ is the point of local minimum of $h'(d)$. There are several
cases, depending on the value of $m^{*}$.
\begin{itemize}
\item If $m^{*}+\mu>h'(\Tld{d}_{1})$, then equation $h'(\Gd)=m^{*}+\mu$ has a
  unique solution $\Gd^{*}$. If $\Gd=\Gd^{*}$ fails (\ref{dconstr}) with
  $m=m^{*}$, then we have polyconvexity with $m=m^{*}$, since
  $F_{1}(\Gd,\Gth)$ has no critical points in $d<\Gth^{2}/4$. If $\Gd=\Gd^{*}$
  satisfies (\ref{dconstr}), then, if $f(\Gd^{*})\ge 0$, then 
  polyconvexity holds with $m=m^{*}$. Here
\begin{equation}
  \label{fofdelta}
f(\Gd)=F_{2}(\Gd,\Gth(\Gd))=h(\Gd)-h(\Gve^{2})-h'(\Gd)(\Gd-\Gve^{2})
-\frac{\Gve^{2}(h'(\Gd)-h'(\Gve^{2}))^{2}}{2\mu}.
\end{equation}
If $f(\Gd^{*})<0$, then we can try to find a better choice for $m<m^{*}$. In
this case, all solutions $\Gd$ of $h'(\Gd)=m+\mu$ will be smaller than
$\Gd^{*}$ and therefore (\ref{dconstr}) will be satisfied for all roots of
$h'(\Gd)=m+\mu$, for any $m\le m_{*}$. Polyconvexity will hold if $f(\Gd)\ge 0$ for all roots of
$h'(\Gd)=m+\mu$ for some choice of $m\le m^{*}$.
\item If $m^{*}+\mu<h'(\Tld{d}_{2})$, then equation $h'(\Gd)=m^{*}+\mu$ has a
  unique solution $\Gd_{*}$. If $\Gd=\Gd_{*}$ fails (\ref{dconstr}) with
  $m=m^{*}$, then we have polyconvexity with $m=m^{*}$, since in the case
  $F_{1}(\Gd,\Gth)$ has no critical points in $d<\Gth^{2}/4$. If $\Gd=\Gd_{*}$
  satisfies (\ref{dconstr}), then, for any $m\le m^{*}$ there will be a unique
  solution $\Gd$ of $h'(\Gd)=m+\mu$, satisfying $\Gd\le\Gd_{*}$. In this case
  polyconvexity fails \IFF $f(\Gd)<0$ for all $\Gd<\Gd_{*}$.
\item If $m^{*}+\mu\in(h'(\Tld{d}_{2}),h'(\Tld{d}_{1}))$, then
  $h(\Gd)=m^{*}+\mu$ has 3 real roots. If even the smallest root $\Gd_{*}$
  does not satisfy (\ref{dconstr}) with $m=m^{*}$, then polyconvexity holds,
  since $F_{1}(\Gd,\Gth)$ has no critical points. If the smallest root
  $\Gd_{*}$ satisfies (\ref{dconstr}), then the smallest
  root of $h(\Gd)=m+\mu$ will satisfy (\ref{dconstr}) for all
  $m\le m^{*}$. Then, if $f(\Gd)<0$ for all $\Gd\le\Gd_{*}$, then polyconvexity
  fails. However, if there are values of $\Gd\le\Gd_{*}$, such that $f(\Gd)\ge
  0$, then it does not imply polyconvexity. For polyconvexity to hold we must
  have $f(\Gd)\ge 0$ for \emph{all} roots of $h'(\Gd)=m+\mu$, which satisfy
  (\ref{dconstr}).
\end{itemize}
More clarity regarding  which case we need to  deal  with can be obtained in the limit
$\mu\to 0$.

\section{Limiting case $\mu\to 0$}
\setcounter{equation}{0}
\label{sec:limcase}
 In the previous section we have derived equations of the
  secondary jump set, conditions for polyconvexity of points $\Gve\BI_{2}$ and
  a differential equation implicitly determining the nucleation bound
  $\Gve_{\infty}\BI_{2}$. In the asymptotic limit $\mu\to 0$ these implicit
  conditions can be made explicit. We emphasize that we consider here the
  family of two-well Hadamard materials with fixed nonlinear potential $h(d)$
  and variable $\mu\to 0$.

\emph{Secondary jump set.} Expanding equation (\ref{jsexpmu}) to first order in $\mu$ we obtain
\begin{equation}
  \label{jsasym}
  \Gve_{+}=\frac{d_{2}}{\Gve_{0}}-\frac{\mu}{4\Gve_{0}^{3}(d_{2}-d_{1})}+O(\mu^{2}),\qquad
\Gve_{-}=\frac{d_{1}}{\Gve_{0}}+\frac{\mu}{4\Gve_{0}^{3}(d_{2}-d_{1})}+O(\mu^{2}).
\end{equation}
Since $d_{1}$ and $d_{2}$ are fixed, we consider the strains $\Gve_{\pm}$ as functions of
$\Gve_{0}$ and $\mu$, even if we suppress this in the notations. Clearly, when
$\mu\to 0$ we have $\Gve_{+}\to d_{2}/\Gve_{0}$, $\Gve_{-}\to
d_{1}/\Gve_{0}$. 

The parametric equations $(x_{0}(\Gve_{0};\mu),y_{0}(\Gve_{0};\mu))$ of
secondary jump set converge, when $\mu\to 0$, to the hyperbola $x_{0}y_{0}=d_{1}$. In
particular, $d_{0}(\Gve_{0},\mu)\to d_{1}$, as $\mu\to 0$. The volume fraction
$\Gl$ of the rank-one laminate used in the second rank laminate is also a
function of $\Gve_{0}$ and $\mu$ and must have a limit (at least along a
subsequence) $\Gl(\Gve_{0};\mu)\to\Gl_{0}(\Gve_{0})$, as $\mu\to 0$. Equation
(\ref{trac}) shows that $d_{0}=d_{1}+\mu\Gd+O(\mu^{2})$, where $\Gd$ solves
\begin{equation}
  \label{tracmu0}
  \frac{\bra{d}}{\Gve_{0}}\left(\Gve_{0}+\frac{1}{\Gve_{0}}
\left(\frac{d_{1}}{d_{2}}{\Gve_{0}^{2}}-\frac{d_{1}+d_{2}}{2\Gve_{0}^{2}}\bra{d}\right)
\right)-d_{1}-2\Gd(d_{2}-d_{1})^{2}\frac{\bra{d}^{2}}{\Gve_{0}^{2}}=0,
\end{equation}
where $\bra{d}=\Gl d_{2}+(1-\Gl)d_{1}$. Equation (\ref{tracmu0}) was obtained
simply by passing to the limit as $\mu\to 0$ in equation (\ref{trac}).

When we pass to the limit as $\mu\to 0$ in (\ref{Maxwell}) we obtain 
\begin{equation}
  \label{Maxwellmu0}
\frac{(\bra{d}-d_{1})^{2}(\Gve_{0}^{4}+\bra{d}^{2}-2d_{2}\bra{d})}
{2\Gve_{0}^{2}\bra{d}^{2}}=0.
\end{equation}
The dependence of $\bra{d}$ on the volume fraction $\Gl$ is essential and
should not disappear in the limit $\mu\to 0$. Therefore, the solution of
(\ref{Maxwellmu0}) that we are after is
\begin{equation}
  \label{dbar}
  \bra{d}=d_{2}-\sqrt{d_{2}^{2}-\Gve_{0}^{4}},
\end{equation}
where the choice of the root was dictated by the requirement that $\bra{d}\le
d_{2}$. Combining this with the requirement that $\bra{d}\ge d_{1}$ we obtain the inequality
\begin{equation}
  \label{epsobds}
  \sqrt[4]{d_{2}^{2}-(d_{2}-d_{1})^{2}}\le\Gve_{0}\le\sqrt{d_{2}}.
\end{equation}
Substituting (\ref{dbar}) into (\ref{tracmu0}) we now can write  the explicit formula for
$\Gd$:
\begin{equation}
  \label{d0asym}
\Gd=\frac{\Gve_{0}^{4}(d_{2}-d_{1})-2(d_{2}^2-\Gve_{0}^4)(d_{2}-\sqrt{d_{2}^2-\Gve_{0}^4})}{4\Gve_{0}^2(d_{2} - d_{1})^2(d_{2}-\sqrt{d_{2}^2-\Gve_{0}^4})^{2}}.
\end{equation}
It looks as if in order to obtain the correct asymptotics of the secondary jump set we need   to compute the
first order asymptotics of $\bra{\Gve}$:
\begin{equation}
  \label{ebarasym}
  \bra{\Gve}=\frac{d_{2}-\sqrt{d_{2}^{2}-\Gve_{0}^{4}}}{\Gve_{0}}+\Tld{\Gve}\mu+O(\mu^{2}).
\end{equation}
In fact, this is not necessary because the leading order asymptotics of $d_{0}$
is \emph{a constant} $d_{1}$. In that case, as far as the first order
asymptotics as $\mu\to 0$ is concerned, using (\ref{ebarasym}) simply corresponds
to reparametrizing the curve
\begin{equation}
  \label{simpsecjs}
  \begin{cases}
    x_{0}=\dfrac{d_{2}-\sqrt{d_{2}^{2}-\Gve_{0}^{4}}}{\Gve_{0}},\\[2ex]
    y_{0}=\dfrac{d_{1}+\mu\Gd(\Gve_{0})}{x_{0}},
  \end{cases}
\end{equation}
 where the range of the parameter $\Gve_{0}$ along the secondary jump set is
given in (\ref{epsobds}).
Indeed, if we change the curve parameter $\Gve_{0}$ to $\Gve_{0}+\mu \Tld{\Gve}/x'_{0}(\Gve_{0})$, then
\[
x_{0}\left(\Gve_{0}+\frac{\mu \Tld{\Gve}}{x'_{0}(\Gve_{0})}\right)=
x_{0}(\Gve_{0})+\mu \Tld{\Gve}+O(\mu^{2}).
\]
At the same time
\[
y_{0}\left(\Gve_{0}+\frac{\mu \Tld{\Gve}}{x'_{0}(\Gve_{0})}\right)=
\frac{d_{1}}{x_{0}(\Gve_{0})}-\frac{\mu d_{1}\Tld{\Gve}}{x_{0}(\Gve_{0})^{2}}
+\frac{\mu\Gd(\Gve_{0})}{x_{0}(\Gve_{0})}+O(\mu^{2})=
\frac{d_{1}+\mu\Gd}{x_{0}+\mu \Tld{\Gve}}+O(\mu^{2}).
\]
We conclude that equation (\ref{simpsecjs}) correctly describes the
asymptotics of the secondary jump set with $O(\mu^{2})$ error, where the
parameter $\Gve_{0}$ varies according to (\ref{epsobds}). When
$\Gve_{0}=\sqrt{d_{2}}$, the secondary jump set passes through one of the
W-points. When $\Gve_{0}=\sqrt[4]{d_{2}^{2}-(d_{2}-d_{1})^{2}}$ it passes
through the limiting point of the ``Legendre-Hadamard for phase boundaries''
bound (see \cite{grtrlhqcx}), that for small $\mu$ lies on the dashed part of
the jump set in Fig.~\ref{fig:secjs}.  The plot of (\ref{simpsecjs}) would be 
indistinguishable from the numerically obtained secondary jump set, if plotted
in Fig.~\ref{fig:secjs}.

\begin{figure}[t]
  \centering
  \includegraphics[scale=0.5]{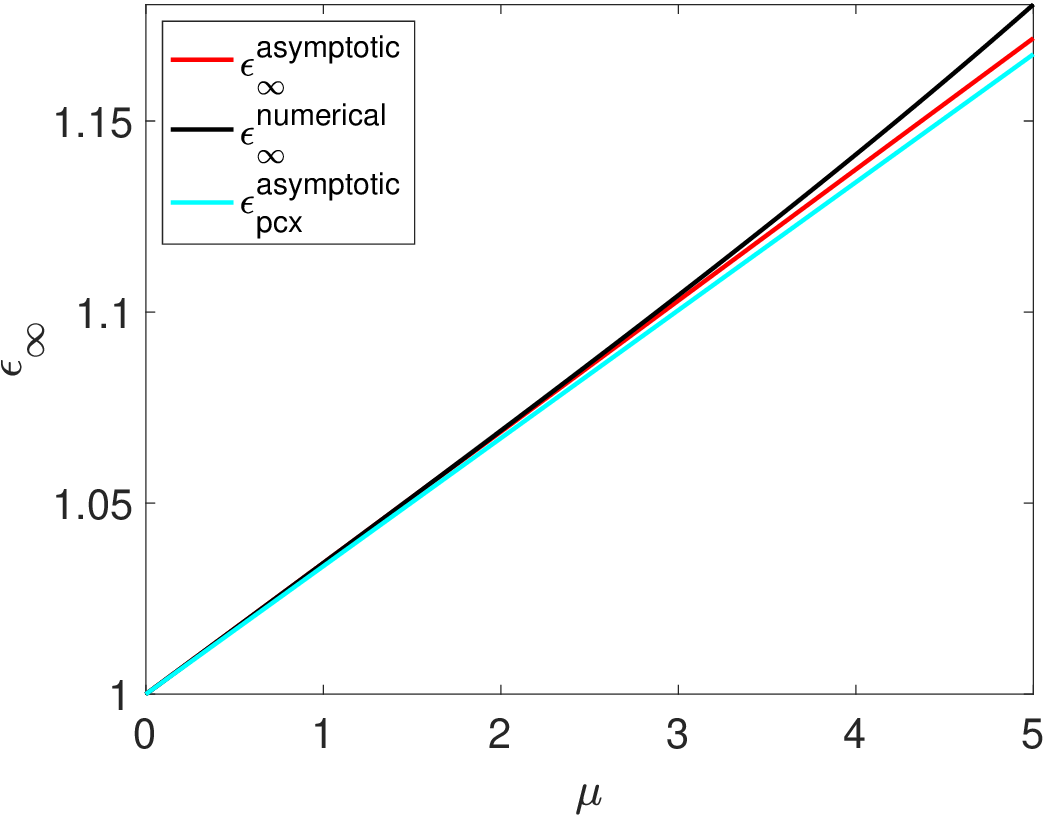}
  \caption{Comparison between the asymptotics (\ref{epsinf}) of
    $\eps_{\infty}$ (denoted here by $\eps_{\infty}^{\rm asymptotic}$)  and
    $\eps_{\infty}^{\rm numerical}$ obtained from the numerical solution of
    (\ref{radequil}). The plot also shows the asymptotics of $\Gve_{\rm pcx}$ (denoted here by $\eps_{\rm pcx}^{\rm asymptotic}$),
 obtained from  (\ref{pcxasym}).}
  \label{fig:hydrobinodal}
\end{figure}
\emph{Circular nucleus. }In the near-liquid limit $\mu\to 0$ we can find the asymptotics of
the solution explicitly. We know that in the limit $\mu\to 0$ the field
$d(\Bx)=\det\Grad\By(\Bx)$ must approach $d_{1}$. Hence,
\[
\frac{\eta\eta'}{r}=d_{1}+\mu\Gd(r)+O(\mu^{2}),\quad r>1.
\]
That implies 
\begin{equation}
  \label{eta}
  \eta(r)=\sqrt{d_{1}r^{2}+c_{0}}+\mu\Tld{\eta}(r)+O(\mu^{2}),
\end{equation}
and therefore,
\[
\Gd(r)=\nth{r}\left(\Tld{\eta}(r)\sqrt{d_{1}r^{2}+c_{0}}\right)'.
\]
Substituting this ansatz into (\ref{radequil}) we obtain
\begin{equation}
  \label{radeqred}
  \mu\frac{\sqrt{d_{1}r^{2}+c_{0}}}{r}h''(d_{1})\Gd'(r)+
\mu\left(\frac{d_{1}r}{\sqrt{d_{1}r^{2}+c_{0}}}+
\frac{\sqrt{d_{1}r^{2}+c_{0}}}{r}\right)'+O(\mu^{2})=0.
\end{equation}
Initial conditions from (\ref{radequil}) imply that
\[
c_{0}=d_{2}-d_{1},\quad\Tld{\eta}(1)=-\frac{d_{2}-d_{1}}{4d_{2}^{3/2}h''(d_{2})},\quad
\Tld{\eta}'(1)=\frac{d_{1}(d_{2}-d_{1})}{2d_{2}^{3/2}}\left(\nth{d_{1}h''(d_{1})}+
\nth{2d_{2}h''(d_{2})}\right).
\]
Equation (\ref{radeqred}) is easy to integrate (observing that
$\sqrt{d_{1}r^{2}+c_{0}}/r$ is decreasing from $\sqrt{d_{2}}$ to
$\sqrt{d_{1}}$ and is therefore uniformly bounded away from zero and
$\infty$).
\begin{equation}
  \label{etatld}
  h''(d_{1})\Tld{\eta}(r)=\frac{c_{1}r^{2}+c_{2}}{\sqrt{d_{1}r^{2}+c_{0}}}-
\frac{r^{2}}{2\sqrt{d_{1}r^{2}+c_{0}}}\ln\frac{\sqrt{d_{1}r^{2}+c_{0}}}{r}.
\end{equation}
From initial conditions for $\Tld{\eta}(r)$ we obtain
\[
c_{1}=\hf\ln\sqrt{d_{2}},\qquad
c_{2}=-\frac{(d_{2}-d_{1})h''(d_{1})}{4d_{2}h''(d_{2})},
\]
and hence
\begin{equation}
  \label{epsinf}
  \BGve_{\infty}=\left(\sqrt{d_{1}}+
\frac{\mu}{2h''(d_{1})\sqrt{d_{1}}}\ln\frac{\sqrt{d_{2}}}{\sqrt{d_{1}}}\right)\BI_{2}
+O(\mu^{2}).
\end{equation}
Figure~\ref{fig:hydrobinodal} shows the quality of the asymptotics for the
entire range of shear moduli $\mu$. The numbers on the $y$-axis indicate that
even for values of $\mu$ that are not particularly small the asymptotics
(\ref{epsinf}) gives a good approximation of the actual value of
$\Gve_{\infty}$. For example, for $\mu=3$ the relative discrepancy is only around 0.1\%.

 \emph{Polyconvexity limits along $\Gve\BI_{2}$.}
If $\mu=0$, then we know that polyconvexity along $\Gve\BI_{2}$ holds whenever
$\Gve\not\in[\sqrt{d_{1}},\sqrt{d_{2}}]$. In this limiting case our analysis
of polyconvexity applied to $\Gve=\sqrt{d_{1}}$ starts with the minimization
problem (\ref{mstar}), which simplifies\footnote{We can assume. \WLOG, that $h(d)\ge
  0$ and $h(d)=0$ only at $d=d_{1}$ and $d=d_{2}$.}:
\begin{equation}
  \label{mstar0}
  m\le\min_{\Gth\in\bb{R}}\frac{h(d_{1}+\Gth\sqrt{d_{1}}+\Gth^{2}/4)}{\Gth^{2}}=0=m^{*}.
\end{equation}
We first observe that in general $\Gth=0$ is not a minimizer. Then there are 3
minimizers:
\[
\Gth=-4\sqrt{d_{1}},\qquad\Gth=\pm 2\sqrt{d_{2}}-2\sqrt{d_{1}}.
\]
 When $\mu$ is positive but small, we examine the
  polyconvexity of points $\Gve=\sqrt{d_{1}}+x$, where $x$ is small. We know
that as $x$ increases, the polyconvexity will fail before $\Gve\BI_{2}$
reaches the point on the secondary jump set, which is known to be
unstable. Hence, we may regard the variable $x$ to be of order $\mu$, which
permits us to compute the asymptotics of all quantities necessary to establish
polyconvexity.

When $x>0$, the minimizer $\Gth(x)$ of (\ref{mstar}) must be located near
one of the above 3 minimizers of (\ref{mstar0}). We can then write $\Gth=\Gth_{0}+y$ for the
minimizer, where $\Gth_{0}$ denotes one of the 3. If we write the function
under the minimum as $H(\Gve,\Gth)$, then at the minimum we must have $\Md
H/\Md\Gth=0$, which gives the equation
\[
x\mix{H}{\Gth}{\Gve}+y\hess{H}{\Gth}=0
\]
relating the infinitesimals $x$ and $y$.
After solving for $y$ and substituting back into $H$ we obtain
\[
H=x\left(\dif{H}{\Gve}-\dif{H}{\Gth}\frac{\mix{H}{\Gth}{\Gve}}{\hess{H}{\Gth}}\right),
\]
where derivatives are evaluated at $(\sqrt{d_{1}},\Gth_{0})$. Maple
calculation yields
\[
H=
\begin{cases}
\frac{x}{2}\sqrt{d_{1}}h''(d_{1}),&\Gth_{0}=-4\sqrt{d_{1}},\\
\frac{xd_{1}h''(d_{1})}{\sqrt{d_{1}}+\sqrt{d_{2}}},&\Gth_{0}=-2\sqrt{d_{2}}-2\sqrt{d_{1}},\\
\frac{xd_{1}h''(d_{1})}{\sqrt{d_{1}}-\sqrt{d_{2}}},&\Gth_{0}=2\sqrt{d_{2}}-2\sqrt{d_{1}}.
\end{cases}
\]
This shows that $\Gth=2\sqrt{d_{2}}-2\sqrt{d_{1}}+y$ is the minimizer, while
\[
m^{*}=\mu-\frac{4xd_{1}}{\sqrt{d_{2}}-\sqrt{d_{1}}}h''(d_{1})+O(x^{2}).
\]
In particular, the equation $h'(\Gd)=m^{*}+\mu$ will have 3 real roots. 
Let us determine how many of them satisfy (\ref{dconstr}), which in the limit
$\mu\to 0$, $x\to 0$ reads
\begin{equation}
  \label{dcosnstr0}
  \Gd\le\left(d_{1}h''(d_{1})\frac{\sqrt{d_{1}}+\sqrt{d_{2}}}{\sqrt{d_{2}}-\sqrt{d_{1}}}\frac{x}{\mu}\right)^{2}.
\end{equation}
\begin{figure}[t]
  \centering
\includegraphics[scale=0.5]{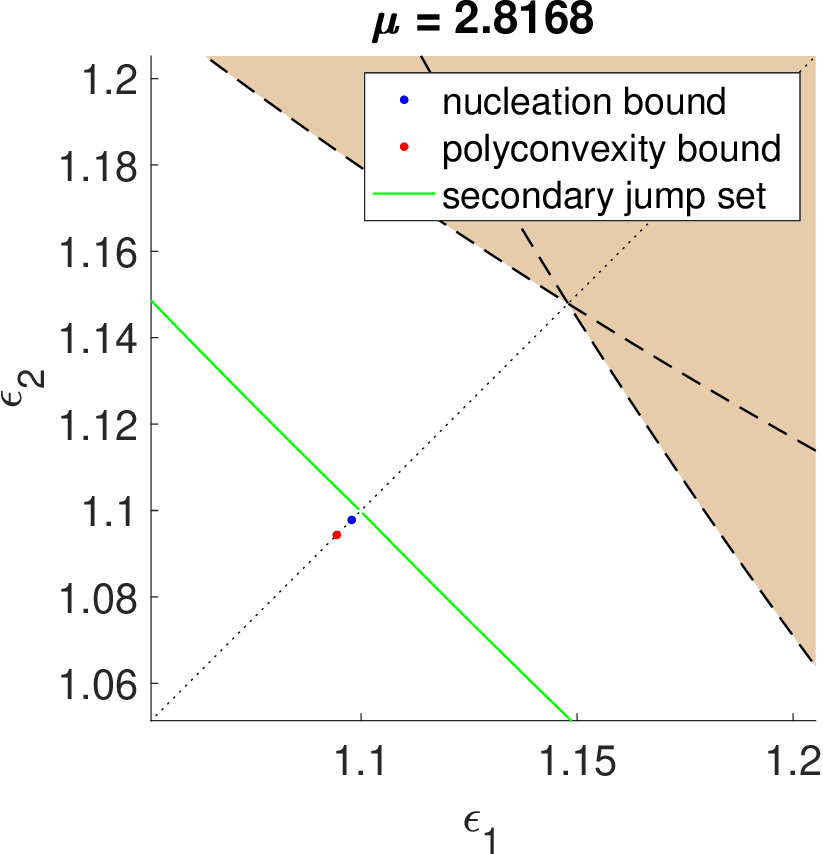}
  \caption{Bounds on the binodal from the inside and the outside of the
    binodal region along hydrostatic strains.}
  \label{fig:binodal1}
\end{figure}
The smallest root $\Gd_{*}$ of $h'(\Gd)=m^{*}+\mu$ has the asymptotics
\[
\Gd_{*}=d_{1}+\frac{\mu+m^{*}}{h''(d_{1})}+O(\mu^{2}+x^{2}).
\]
Therefore, it will fail (\ref{dcosnstr0}), when 
\begin{equation}
  \label{xasym}
x\le\frac{\mu}{h''(d_{1})\sqrt{d_{1}}}\frac{\sqrt{d_{2}}-\sqrt{d_{1}}}{\sqrt{d_{2}}+\sqrt{d_{1}}}
+O(\mu^{2}).
\end{equation}
If $x$ is larger than the upper bound (\ref{xasym}), then $\Gd_{*}$ satisfies
(\ref{dcosnstr0}). 

Let us now show that in this case $f(\Gd)<0$ for all
$\Gd\le\Gd_{*}$ for sufficiently small $\mu$. Indeed, $h'(\Gd)$ is a monotone
increasing function on $\Gd\le\Gd_{*}$. Therefore, either $\Gd$ is close to
$d_{1}$ or $h'(\Gd)-h'(\Gve^{2})$ is not small. In the latter case $f(\Gd)$,
given by (\ref{fofdelta}) is clearly negative, as its last term tends to
$-\infty$, when $\mu\to 0$. When $\Gd$ is close to $d_{1}$, then, using the
Taylor expansion of $h(\Gd)$ centered at $\Gve^{2}$, we obtain
\[
f(\Gd)=-\hf h''(\Gve^{2})(\Gd-\Gve^{2})^{2}-\frac{\Gve^{2}h''(\Gve^{2})^{2}(\Gd-\Gve^{2})^{2}}{2\mu}+O\left(\frac{(\Gd-\Gve^{2})^{3}}{\mu}\right),
\] 
which is obviously negative, when $\mu$ is sufficiently small. We conclude
that polyconvexity at $\Gve\BI_{2}$ holds when $\Gve\le\Gve_{\rm pcx}$, where
\begin{equation}
  \label{pcxasym}
\Gve_{\rm pcx}=\sqrt{d_{1}}+\frac{\mu}{h''(d_{1})\sqrt{d_{1}}}\frac{\sqrt{d_{2}}-\sqrt{d_{1}}}{\sqrt{d_{2}}+\sqrt{d_{1}}}+O(\mu^{2}).
\end{equation}
Our Fig.~\ref{fig:binodal1}, where $\Gve_{\rm pcx}\BI_{2}$ is represented by the red dot,
shows that $\Gve_{\infty}\BI_{2}$ fails to be polyconvex, but by a very slim
margin.  Our numerical investigations (to be reported
  elsewhere) shows that the  ordering of the bounds in Fig.~\ref{fig:binodal1} persists on
the entire range of $\mu$.  In Fig.~\ref{fig:binodal1} we see that the remaining gap between established stability
(along the bisector below the red dot) and established instability (along the
bisector above the blue dot) is very small.

\section{A glimpse into the relaxed energy}
\setcounter{equation}{0}
\label{sec:hyp}
\emph{Hypothetical bounds on the binodal.} 
We have seen in the foregoing discussion that the energy  $W(\BF)$ is  not polyconvex  at
$\BF=\Gve_{\infty}\BI_{2}$.  This is not very surprising, since polyconvexity
is usually strictly stronger that quasiconvexity and we expect and conjecture
that $\BF=\Gve_{\infty}\BI_{2}$ lies on the binodal---at the very edge of
quasiconvexity. 

First, we recall our observation that if $\BF=\Gve_{\infty}\BI_{2}$ is stable,
then for every $|\Bx|>1$ the deformation gradients
\[
\Grad\By(\Bx)=\eta'(r)\tns{\hat{\Bx}}+\frac{\eta(r)}{r}(\BI_{2}-\tns{\hat{\Bx}})
\]
are also  stable in the sense of Definition~\ref{def:stabF}. This observation
would  then provide a bound on the whole binodal from the outside. 

\begin{figure}[h!]
  \centering
  \includegraphics[scale=0.5]{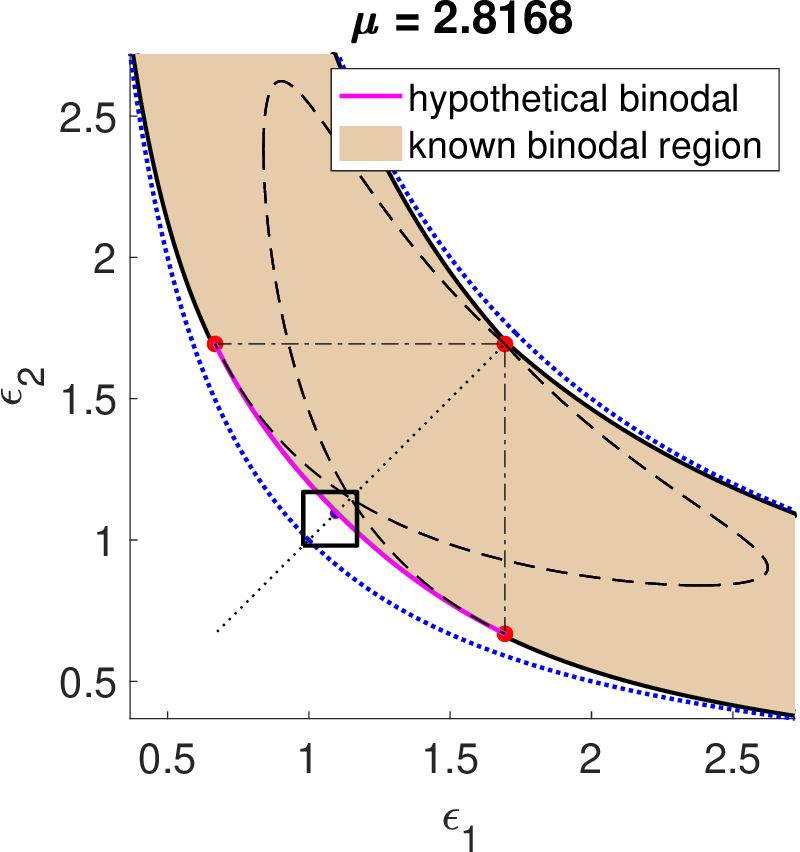}~~~~~~~
  \includegraphics[scale=0.5]{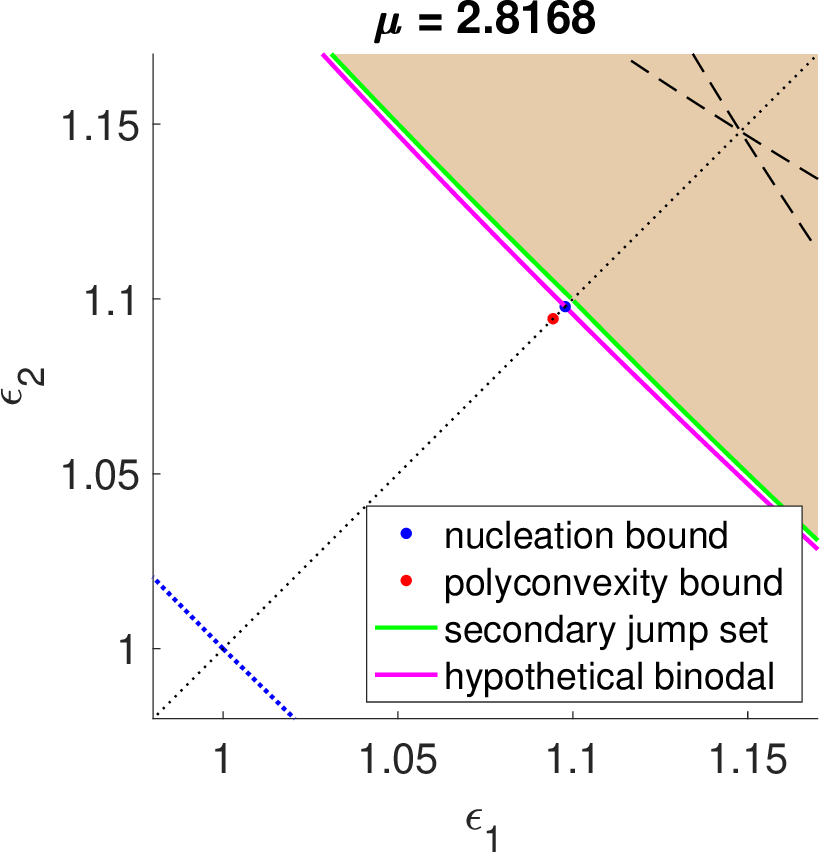}
  \caption{A hypothetical bound on the binodal region from the outside under the condition of stability of $\Gve_{\infty}\BI_{2}$. The right panel is the blow-up of the 
square in the left panel.}
  \label{fig:hyp}
\end{figure}

Note next that for the entire range of $\mu$ for which W-points
are polyconvex the union of the curves
\begin{equation}
  \label{binbd}
  \begin{cases}
      \Gve_{1}=\frac{\eta(r)}{r},\\
  \Gve_{2}=\eta'(r),
  \end{cases}\text{ and }
\begin{cases}
      \Gve_{1}=\eta'(r)\\
  \Gve_{2}=\frac{\eta(r)}{r},
  \end{cases}\quad r>1
\end{equation}
appear as almost indistinguishable from the secondary jump set curves shown in
green in Fig.~\ref{fig:secjs}. This is more clear in Fig.~\ref{fig:hyp}
showing the blown-up part of the strain space from Fig.~\ref{fig:binodal1},
where the curves (\ref{binbd}) shown in magenta are meeting at the blue point
from Fig.~\ref{fig:binodal1} entering it with slope $-1$. Assuming the
conjectured stability of $\Gve_{\infty}\BI_{2}$, the magenta curve must lie
outside of the binodal region, while secondary jump set lies in its
interior.
Thus, the binodal of the energy (\ref{enerex}) would have to lie between the
green and the magenta curves. We conjecture that the magenta curve is in fact
the actual binodal of the energy (\ref{enerex}).  Independently of whether
this more general conjecture is true, the magenta line represents a rather
tight outside bound on the binodal region which hinges only on a more modest
assumption of the stability of $\Gve_{\infty}\BI_{2}$.

\begin{figure}[t]
  \centering
  \includegraphics[scale=0.4]{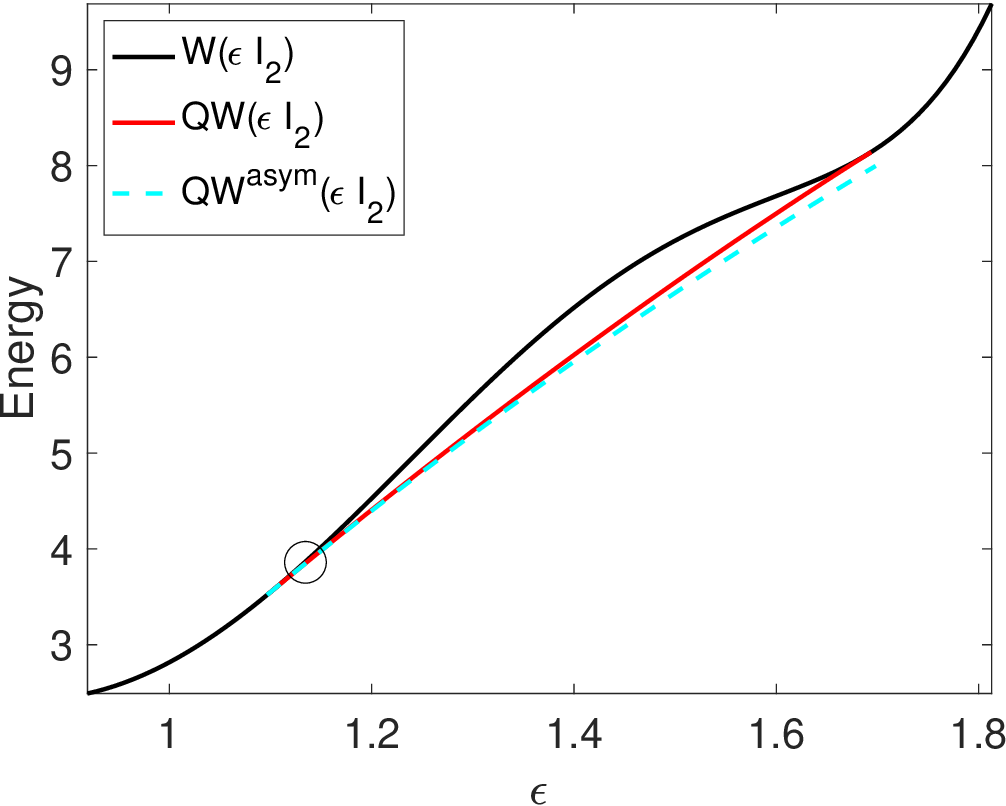}~~~~~~~~~
\includegraphics[scale=0.4]{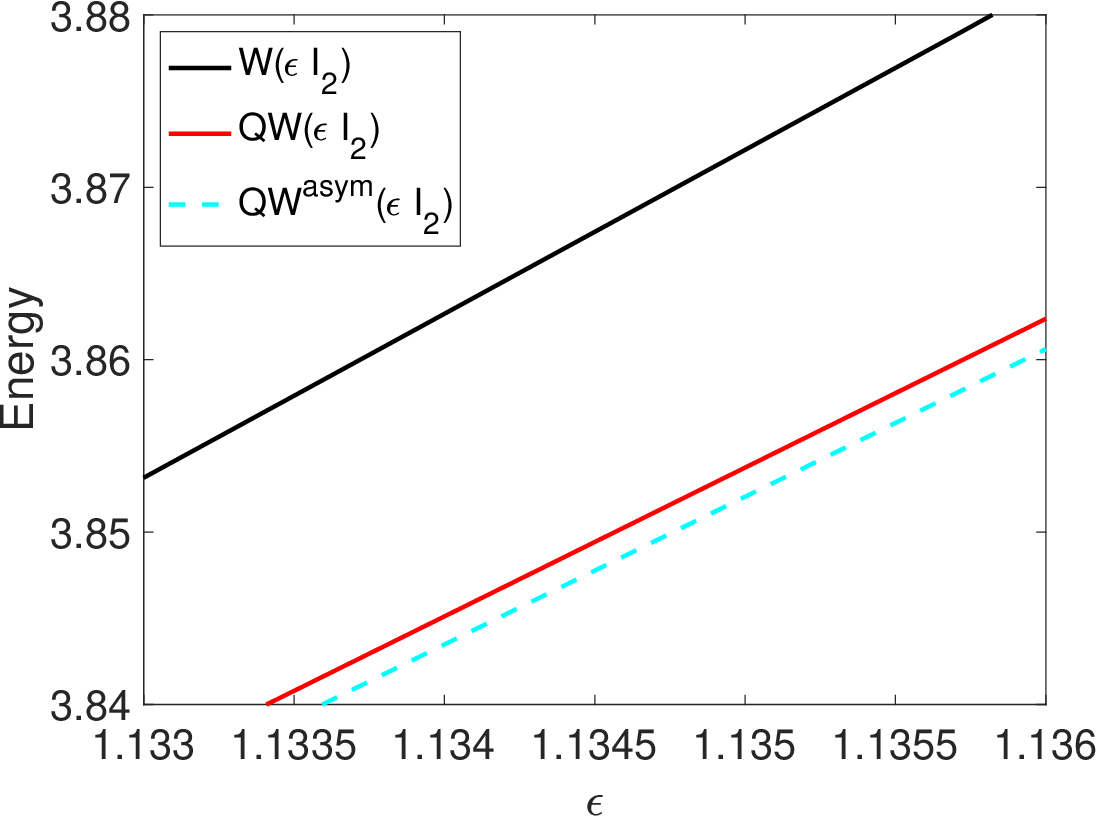}
  \caption{Quasiconvex envelope of $W(\BF)$ restricted to hydrostatic strains
    $\BF=\Gve\BI_2$. The right panel shows the blow-up of a subset of the small circle in the left panel.}
  \label{fig:QW}
\end{figure}

Another byproduct of the assumed stability of $\Gve_{\infty}\BI_{2}$ would be
the formula for the quasiconvex envelope $QW(\BF)$ for hydrostatic strains
$\BF$.  If $\BF=\Gve_{\infty}\BI_{2}$ is stable, then our radial solution
$\Grad\By(\Bx)=\eta(r)\hat{\Bx}$ of (\ref{radequil}) is also a global
minimizer in every finite ball $B(0,R)$, where it satisfies the affine
boundary condition $\By(\Bx)=(\eta(R)/R)\Bx$, $\Bx\in\Md B(0,R)$
\cite{grtrhard}. The energy of such configurations must necessarily be
$QW(\eta(R)\BI_{2}/R)|B(0,R)|$. This permits us to compute $QW(\Gve\BI_{2})$
for all $\Gve$, as the energy of configurations $\By(\Bx)=\eta(r)\hat{\Bx}$ in
$B(0,R)$. Using the Clapeyron-type formula for the nonlinear elastic energy
stored in an equilibrium stationary configuration we obtain for $\BF=\eta(R)\BI_{2}/R$:
\cite{grtrhard}.
\begin{equation}
  \label{QWform}
  |B(0,R)|QW(\BF)=\hf\int_{\Md B(0,R)}\{\BP(\Grad\By)\Bn\cdot\By+\BP^{*}(\Grad\By)\Bn\cdot\Bx\}dS.
\end{equation}
Finally, substituting $\Bn=\hat{\Bx}$, $\By=\eta(r)\hat{\Bx}$ into (\ref{QWform}) we
obtain
\begin{equation}
  \label{QWexpl}
  QW\left(\frac{\eta(R)}{R}\BI_{2}\right)=2(\mu-h'(d))d
-\mu\eta'(R)^{2}+(2h'(d)+\mu)\frac{\eta(R)^{2}}{R^{2}}+2h(d),
\end{equation}
where
\[
d=\frac{\eta'(R)\eta(R)}{R}.
\]
When $\mu$ is small we can use the asymptotic formulas (\ref{eta}), (\ref{etatld}) 
for $\eta(r)$ to obtain explicit asymptotics $QW^{\rm asym}(\Gve\BI_{2})$ 
for $QW(\Gve\BI_{2})$.  The plot of $QW(\Gve\BI_{2})$, coming from the
numerical solution of (\ref{radequil}), as well as its explicit asymptotic
approximation $QW^{\rm asym}(\Gve\BI_{2})$, superposed on the plot of
$W(\Gve\BI_{2})$ are shown in Fig.~\ref{fig:QW}. 

\section{Conclusions}
\setcounter{equation}{0}
\label{sec:conc}
In this paper we posed the problem  of solving analytically the relaxation
problem for the double well Hadamard energy \eqref{enerex} in two space
dimensions in the limit when the rigidity measure $\mu$ is sufficiently small. However, we  only succeeded in   attaining  a much more modest  goal of  locating a substantial  part of the corresponding binodal region in  the strain space. 

To deal analytically with these challenging questions, we used some of our previously developed methods centered around the computation of the jump set and the identification  of its stable part. While our general methods apply for Hadamard materials in the entire parameter range and are amenable to numerical implementation, in this paper we have chosen to focus only on  explicit asymptotic study of  the ``near-liquid'' regime.

In particular, we managed to show that in  this limit, a subset of the jump set adjacent to the high strain phase
remains stable which ensures that simple lamination delivers the corresponding
part of the binodal. This means that even when the parameter $\mu$ is infinitesimally 
small, the high strain phase maintains its tangential
rigidity at the level which ensures solid-solid like nature of the incipient
phase transition.

By contrast, our analysis showed that the subset of the jump set adjacent to
the low strain and low energy phase is unstable in the $\mu \to 0$
limit. Moreover, the secondary jump set is also unstable in this limit. This
result implies that laminates of any \emph{finite} rank are unstable and
cannot be associated to any part of the binodal in that regime.

Whether the revealed asymmetry of the transformation mechanism between the
direct and reverse transformation is a peculiarity of the Hadamard material or
whether this striking phenomenon has a more general nature, remains to be
established. It shows, however, the intricate role of rigidity in structural
transformations which, even if weak, can produce   complex  microstructural  morphologies  underlying the relaxed energy. This complexity shows that, rather remarkably,  the elastic long-range interactions remain relevant even  when the system is arbitrarily close to the liquid regime. In other words, the disappearing rigidity can be viewed as  the microstructure selection mechanism for the Hadamard liquid  which otherwise comes with an infinite repertoire of possible accommodation mechanisms all having  zero energetic cost.

In order to reconcile the solid-like features of the behavior of the near-liquid material with the behavior of its purely ``liquid'' limit
one can turn from the study of global minima of the energy to the study of almost-minimizers whose energy is only slightly above the energy of the ground state. In this case   the richness of the   repertoire of purely liquid microstructures  corresponding to    $\mu=0$ can be recovered  in the form of such  almost-minimizers of the Hadamard solid with  $\mu$  sufficiently small but finite. 
The physical merit of these projections must be weighed
against other factors that have been neglected in our study, such as surface
energy, crystal anisotropy, spatial  inhomogeneity  and dynamics.

\medskip

\textbf{Funding.} YG was supported by the
National Science Foundation under Grant No. DMS-2005538. The work of LT was supported by the French   grant ANR-10-IDEX-0001-02 PSL.

\textbf{Author contributions.}  Both authors wrote the main manuscript text
and reviewed the manuscript.

\section{Declarations}
\textbf{Competing interests.} The authors declare no competing interests.

\appendix
\section{Calculation of the jump set for Hadamard materials}
\setcounter{equation}{0}
\label{sec:genodgen}
Here we recall the calculation of the jump set from \cite{grtrsolid} for energies  (\ref{enerex}).

We start with the first equation in (\ref{jsgen}) expressing the
kinematic compatibility of the deformation gradients $\BF_{+}$ and
$\BF_{-}$. Taking the determinant of both sides we obtain
\begin{equation}
  \label{dplus}
   d_{+}=d_{-}+\cof\BF_{-}\Bn\cdot\Ba,\quad d_{\pm}=\det\BF_{\pm}.
\end{equation}
Using the formula
\[
\BP(\BF)=\mu\BF+h'(\det\BF)\cof\BF
\]
for the Piola-Kirchhoff stress we compute
\[
\jump{\BP}\Bn=\mu\Ba+\jump{h'}\cof\BF_{-}\Bn,
\]
where we have used the well-known relation
$\cof(\BF_{-}+\Ba\otimes\Bn)\Bn=(\cof\BF_{-})\Bn$. Similarly,
\[
\jump{\BP}^{T}\Ba=\mu|\Ba|^{2}\Bn+\jump{h'}\cof\BF_{-}^{T}\Ba.
\]
Thus, the second and the third equations in (\ref{jsgen}) become
\begin{equation}
  \label{an}
\Ba=-\frac{\jump{h'}}{\mu}\cof\BF_{-}\Bn,\qquad\jump{h'}^{2}\cof(\BC_{-})\Bn=\mu^{2}|\Ba|^{2}\Bn,
\end{equation}
where $\BC_{\pm}=\BF_{\pm}^{T}\BF_{\pm}$ is the Cauchy-Green strain tensor.
We conclude that $\Bn$ must be an eigenvector of $\BC_{-}$. Equations (\ref{an})
permit us to find a relation between the two Cauchy-Green tensors
$\BC_{\pm}$.  Using the kinematic compatibility
equation (\ref{jsgen})$_{1}$ we compute
\[
\BC_{+}=\BC_{-}+\BF_{-}^{T}\Ba\otimes\Bn+\Bn\otimes\BF_{-}^{T}\Ba+|\Ba|^{2}\tns{\Bn}.
\]
Applying $\BF_{-}^{T}$ to the first equation in (\ref{an}) we obtain
$\BF_{-}^{T}\Ba=-(\jump{h'}/\mu)d_{-}\Bn$, so that
\begin{equation}
  \label{jumpC}
  \jump{\BC}=\left(|\Ba|^{2}-\frac{2\jump{h'}d_{-}}{\mu}\right)\tns{\Bn}.
\end{equation}
It follows that the Cauchy-Green tensors $\BC_{+}$ and $\BC_{-}$ are
simultaneously diagonalizable, since, by (\ref{an}) $\Bn$ is an
eigenvector of $\BC_{-}$. According to equation (\ref{jumpC})
symmetric matrices $\BC_{+}$ and $\BC_{-}$ have the same pair of
mutually orthogonal eigenvectors $\Bn$ and $\Bn^{\perp}$ with the same eigenvalues
corresponding to $\Bn^{\perp}$. Hence, singular values of
$\BF_{\pm}$ would be
$(\Gve_{\pm},\Gve_{0})$, the first one corresponding
to the eigenvector $\Bn$ of $\BC_{\pm}$. Substituting the first
equation in (\ref{an}) into (\ref{dplus}) we obtain
\[
d_{+}=d_{-}-\frac{\jump{h'}}{\mu}\cof\BC_{-}\Bn\cdot\Bn=d_{-}-\frac{\jump{h'}d_{-}^{2}}{\mu\Gve_{-}^{2}},
\]
which can be written in the more symmetric form as
\begin{equation}
  \label{jumpS}
  \mu\frac{\jump{d}}{\jump{h'}}=-\Gve_{0}^{2}=-\frac{d_{\pm}^{2}}{\Gve_{\pm}^{2}}.
\end{equation}
This will be the equation for the jump set, when we determine $d_{+}$ as a
function of $d_{-}$ from the Maxwell relation (the last equation in
(\ref{jsgen}), which hasn't been used so far). It is well-known that the
Maxwell relation does not change if we add any quadratic function in $\BF$ to
the energy. Thus, the term $\mu|\BF|^{2}/2$ can be disregarded and the Maxwell
relation becomes
\[
\jump{h}=\lump{h'\cof\BF}\Bn\cdot\Ba.
\]
Recalling that due to (\ref{dplus}) $(\cof\BF_{+})\Bn\cdot\Ba=(\cof\BF_{-})\Bn\cdot\Ba=\jump{d}$ we obtain
\begin{equation}
  \label{Hadmax}
  \jump{h}=\lump{h'}\jump{d}.
\end{equation}
Equation (\ref{Hadmax}) has a geometric meaning. It says that the
secant line joining $(d_{-},h'(d_{-}))$ and $(d_{-},h'(d_{-}))$
together with the graph of $h'(d)$ bound two regions of equal areas.
For a double-well shaped potential $h(d)$ there exists a single
interval $(d_{1},d_{2})$ on which $h(d)$ differs from its convex hull,
which on $(d_{1},d_{2})$ agrees with the common tangent line at
$d_{1}$ and $d_{2}$ to the graph of $h(d)$. In terms of $h'(d)$ this
double-tangency can also be interpreted geometrically as the
horizontal ``Maxwell line'' with the equal area property. In that case
there exist $d_{0}\in(d_{1},d_{2})$, such that for any
$d_{-}\in(d_{1},d_{0})$ there is a unique $d_{+}\in(d_{0},d_{2})$
satisfying (\ref{Hadmax}). Moreover, for every $d_{-}\in(d_{1},d_{0})$
there is a unique $d_{+}=D(d_{-})$ with equal area property. (By
continuity we can set $D(d_{0})=d_{0}$.) Regarding the function $D(d)$
as known, equation (\ref{jumpS}) provides the explicit description of
the jump set in terms of the singular values of $\BF_{\pm}$.

\def\cprime{$'$} \ifx \cedla \undefined \let \cedla = \c \fi\ifx \cyr
  \undefined \let \cyr = \relax \fi\ifx \cprime \undefined \def \cprime
  {$\mathsurround=0pt '$}\fi\ifx \prime \undefined \def \prime {'}
  \fi\def\Ya{Ya}


\begin{thebibliography}{10}

\bibitem{ak}
G.~Allaire and R.~V. Kohn.
\newblock Explicit optimal bounds on the elastic energy of a two-phase
  composite in two space dimensions.
\newblock {\em Quart. Appl. Math.}, LI(4):675--699, December 1993.

\bibitem{achf16}
Mikhail~A. Antimonov, Andrej Cherkaev, and Alexander~B. Freidin.
\newblock Phase transformations surfaces and exact energy lower bounds.
\newblock {\em International Journal of Engineering Science}, 90:153--182,
  2016.

\bibitem{ball10}
J.~M. Ball.
\newblock Progress and puzzles in nonlinear elasticity.
\newblock In J{\"o}rg Schr{\"o}der and Patrizio Neff, editors, {\em Poly-,
  Quasi- and Rank-One Convexity in Applied Mechanics}, pages 1--15. Springer
  Vienna, Vienna, 2010.

\bibitem{bamu84}
J.~M. Ball and F.~Murat.
\newblock {$W\sp{1,p}$}-quasiconvexity and variational problems for multiple
  integrals.
\newblock {\em J. Funct. Anal.}, 58(3):225--253, 1984.

\bibitem{baja15}
J.M. Ball and R.D. James.
\newblock Incompatible sets of gradients and metastability.
\newblock {\em Archive for Rational Mechanics and Analysis}, 218(3):1363--1416,
  2015.

\bibitem{ball02}
John~M. Ball.
\newblock Some open problems in elasticity.
\newblock In {\em Geometry, mechanics, and dynamics}, pages 3--59. Springer,
  New York, 2002.

\bibitem{blar}
D.~M. Barnett, J.~K. Lee, H.~I. Aaronson, and K.~C. Russel.
\newblock The strain energy of a coherent ellipsoidal precipitate.
\newblock {\em Scripta Metall.}, 8:1447--1450, 1974.

\bibitem{chlw95}
Paul~M Chaikin, Tom~C Lubensky, and Thomas~A Witten.
\newblock {\em Principles of condensed matter physics}, volume~10.
\newblock Cambridge university press Cambridge, 1995.

\bibitem{chbh07}
Isaac Chenchiah and Kaushik Bhattacharya.
\newblock The relaxation of two-well energies with possibly unequal moduli.
\newblock {\em Arch. Rat. Mech. Anal.}, 187(3):409--479, 2008.

\bibitem{ciar21}
Philippe~G Ciarlet.
\newblock {\em Mathematical elasticity: Three-dimensional elasticity}.
\newblock SIAM, 2021.

\bibitem{daco81}
B.~Dacorogna.
\newblock A relaxation theorem and its application to the equilibrium of gases.
\newblock {\em Arch. Rational Mech. Anal.}, 77(4):359--386, 1981.

\bibitem{daco82}
B.~Dacorogna.
\newblock Quasiconvexity and relaxation of nonconvex problems in the calculus
  of variations.
\newblock {\em J. Funct. Anal.}, 46(1):102--118, 1982.

\bibitem{Dak08}
B.~Dacorogna.
\newblock {\em Direct methods in the calculus of variations}.
\newblock Springer-Verlag, New York, 2nd edition, 2008.

\bibitem{dcbunv16}
Michelle~M Driscoll, Bryan Gin-ge Chen, Thomas~H Beuman, Stephan Ulrich,
  Sidney~R Nagel, and Vincenzo Vitelli.
\newblock The role of rigidity in controlling material failure.
\newblock {\em Proceedings of the National Academy of Sciences},
  113(39):10813--10817, 2016.

\bibitem{Ericksen:1980:SPT}
J.~Ericksen.
\newblock Some phase transitions in crystals.
\newblock {\em Archive for Rational Mechanics and Analysis}, 73:99--124, 1980.

\bibitem{erick75}
J.~L. Ericksen.
\newblock Equilibrium of bars.
\newblock {\em J. Elasticity}, 5(3--4):191--201, 1975.

\bibitem{erick87}
J.~L. Ericksen.
\newblock Twinning of crystals. {I}.
\newblock In {\em Metastability and incompletely posed problems (Minneapolis,
  Minn., 1985)}, pages 77--93. Springer, New York, 1987.

\bibitem{Ericksen:1991:KCC}
J.~L. Ericksen.
\newblock On kinematic conditions of compatibility.
\newblock {\em Journal of Elasticity}, 26(1):65--74, 1991.

\bibitem{erick92a}
J.~L. Ericksen.
\newblock Bifurcation and martensitic transformations in {B}ravais lattices.
\newblock {\em J. Elasticity}, 28(1):55--78, 1992.

\bibitem{gibbs1876}
J.~Gibbs, Willard.
\newblock On the equilibrium of heterogeneous substances.
\newblock {\em Transactions of the Connecticut Academy}, III:108--248 and
  343--524, 1873 and 1874.

\bibitem{golu89}
Leonardo Golubovi{\'c} and T.~C. Lubensky.
\newblock Nonlinear elasticity of amorphous solids.
\newblock {\em Physical review letters}, 63(10):1082--1085, 1989.

\bibitem{gra}
Y.~Grabovsky.
\newblock Bounds and extremal microstructures for two-component composites: {A}
  unified treatment based on the translation method.
\newblock {\em Proc. Roy. Soc. London, Series A.}, 452(1947):945--952, 1996.

\bibitem{grtrpe}
Y.~Grabovsky and L.~Truskinovsky.
\newblock Roughening instability of broken extremals.
\newblock {\em Arch. Rat. Mech. Anal.}, 200(1):183--202, 2011.

\bibitem{grtrmms}
Y.~Grabovsky and L.~Truskinovsky.
\newblock Marginal material stability.
\newblock {\em Journal of Nonlinear Science}, 23(5):891--969, 2013.

\bibitem{grtrnc}
Yury Grabovsky and Lev Truskinovsky.
\newblock Normality condition in elasticity.
\newblock {\em Journal of Nonlinear Science}, 24(6):1125--1146, 2014.

\bibitem{grtrlhqcx}
Yury Grabovsky and Lev Truskinovsky.
\newblock Legendre-{H}adamard conditions for two-phase configurations.
\newblock {\em Journal of Elasticity}, 123(2):225--243, 2016.

\bibitem{grtrsolid}
Yury Grabovsky and Lev Truskinovsky.
\newblock Explicit relaxation of a two-well hadamard energy.
\newblock {\em Journal of Elasticity}, 135(1-2):351--373, 2019.

\bibitem{grtrpcx}
Yury Grabovsky and Lev Truskinovsky.
\newblock When rank-one convexity meets polyconvexity: An algebraic approach to
  elastic binodal.
\newblock {\em J. Nonlinear Sci.}, 28(1):229--253, 2019.

\bibitem{grtrliquid}
Yury Grabovsky and Lev Truskinovsky.
\newblock Ubiquity of infinite rank laminates.
\newblock {\em to be submitted}, In preparation.

\bibitem{grtrhard}
Yury Grabovsky and Lev Truskinovsky.
\newblock A vectorail elasticity problem with many global but no local
  minimizers.
\newblock {\em J. Elasticity}, to appear.

\bibitem{hada03}
J.~Hadamard.
\newblock {\em Le{\c{c}}ons sur la propagation des ondes et les {\'e}quations
  de l'hydrodynamique.}
\newblock Hermann, Paris, 1903.

\bibitem{john66}
Fritz John.
\newblock Plane elastic waves of finite amplitude. hadamard materials and
  harmonic materials.
\newblock {\em Communications on Pure and Applied Mathematics}, 19(3):309--341,
  1966.

\bibitem{karo72}
V.~Kardonski and Roitburd.
\newblock On the shape of coherent precipitates.
\newblock {\em Phys. Met. Metallurg. USSR}, 33:210--212, 1972.

\bibitem{kh}
A.~G. Khachaturyan.
\newblock {\em Theory of structural transformation in solids}.
\newblock Wiley, New York, 1983.

\bibitem{khach67}
Armen~G Khachaturyan.
\newblock Some questions concerning the theory of phase transformations in
  solids.
\newblock {\em Soviet Phys. Solid State}, 8(9):2163--2168, 1967.

\bibitem{DW}
R.~V. Kohn.
\newblock The relaxation of a double-well energy.
\newblock {\em Continuum Mech. Thermodyn.}, 3:193--236, 1991.

\bibitem{krro19}
Martin Kru{\v{z}}{\'\i}k and Tom{\'a}{\v{s}} Roub{\'\i}{\v{c}}ek.
\newblock {\em Mathematical methods in continuum mechanics of solids}.
\newblock Springer, 2019.

\bibitem{kufr88}
L.~B. Kublanov and A.~B. Freidin.
\newblock Nuclei of a solid phase in a deformable material.
\newblock {\em Prikl. Mat. Mekh.}, 52(3):493--501, 1988.

\bibitem{lali13v5}
Lev~Davidovich Landau and Evgenii~Mikhailovich Lifshitz.
\newblock {\em Statistical Physics: Volume 5}, volume~5.
\newblock Elsevier, 2013.

\bibitem{lba}
J.~K. Lee, D.~M. Barnett, and H.~I. Aaronson.
\newblock The elastic strain energy of coherent ellipsoidal precipitates in
  anisotropic crystalline solids.
\newblock {\em Metall. Trans. A}, 8A:963--970, 1977.

\bibitem{maxwell1875}
J.~C. Maxwell.
\newblock On the dynamic evidence of the molecular composition of bodies.
\newblock {\em Nature}, 11(279-280):357--359, 374--377, 1875.

\bibitem{pineau}
A.~Pineau.
\newblock Influence of uniaxial stress on the morphology of coherent
  precipitates during coarsening --- elastic energy considerations.
\newblock {\em Acta Metall.}, 24:559--564, 1976.

\bibitem{pipk91}
Allen~C Pipkin.
\newblock Elastic materials with two preferred states.
\newblock {\em The Quarterly Journal of Mechanics and Applied Mathematics},
  44(1):1--15, 1991.

\bibitem{silh13}
Miroslav Silhavy.
\newblock {\em The mechanics and thermodynamics of continuous media}.
\newblock Springer Science \& Business Media, 2013.

\bibitem{vandW03}
J.D. van~der Waals.
\newblock The equilibrium between a solid body and a fluid phase, especially in
  the neighbourhood of the critical state.
\newblock In {\em KNAW, Proceedings}, volume~6, pages 1903--1904, 1903.

\end{thebibliography}
\end{document}